%% file: main.tex
\documentclass[preprint,12pt]{elsarticle}
\usepackage{xcolor}
\usepackage{tcolorbox}
\usepackage{graphicx}
\usepackage{multirow}
\usepackage{microtype}
\usepackage{url}

\usepackage{amssymb}
\usepackage{enumerate}

\begin{document}

\begin{frontmatter}



\title{
Insights into Software Development Approaches: Mining Q\&A Repositories}


\author[inst1]{Arif Ali Khan}

\affiliation[inst1]{Organization={M3S Empirical Software Engineering Research Unit, University of Oulu,},Department and Organization
            city={Oulu},
            postcode={FI-90014}, 
            country={Finland}}
           
\author[inst2]{Javed Ali Khan}
\affiliation[inst2]{organization={Department of Software Engineering,University of Science and Technology},
            city={Bannu},
            postcode={28100}, 
            country={Pakistan}}
            
\author[inst3]{Muhammad Azeem Akbar}
\affiliation[inst3]{organization={Software Engineering Department,
Lappeenranta-Lahti University of Technology},
            city={Lappeenranta},
            postcode={15210}, 
            country={Finland}}
            
\author[inst4]{Peng Zhou}
\affiliation[inst4]{organization={College of Computer Science and Technology, Nanjing University of Aeronautics and Astronautics},
            city={Nanjing},
            postcode={210016}, 
            country={China}}
            
\author[inst5]{Mahdi Fahmideh}
\affiliation[inst5]{organization={School of Business, University of Southern
Queensland},
            city={Toowoomba},
            postcode={QLD 4350}, 
            country={Australia}}

\begin{abstract}
\textbf{Context:} Software practitioners adopt approaches like DevOps, Scrum, and Waterfall for high-quality software development. However, limited research has been conducted on exploring software development approaches concerning practitioners' discussions on Q\&A forums. \textbf{Objective:} We conducted an empirical study to analyze developers' discussions on Q\&A forums to gain insights into software development approaches in practice. \textbf{Method:} We analyzed 13,903 developers' posts across Stack Overflow (SO), Software Engineering Stack Exchange (SESE), and Project Management Stack Exchange (PMSE) forums. A mixed method approach, consisting of the topic modeling technique (i.e., Latent Dirichlet Allocation (LDA)) and qualitative analysis, is used to identify frequently discussed topics of software development approaches, trends (popular, difficult topics), and the challenges faced by practitioners in adopting different software development approaches. \textbf{Findings:} We identified 15 frequently mentioned software development approaches topics on Q\&A sites and observed an increase in trends for the top-3 most difficult topics requiring more attention. Finally, our study identified 49 challenges faced by practitioners while deploying various software development approaches, and we subsequently created a thematic map to represent these findings.
\textbf{Conclusions:} The study findings serve as a useful resource for practitioners to overcome challenges, stay informed about current trends, and ultimately improve the quality of software products they develop.
\end{abstract}




\end{frontmatter}


\input{introduction.tex}

\input{motivation.tex}
\input{related-work.tex}

\input{research-method.tex}

\input{results.tex}
\input{implication.tex}

\input{threats.tex}
\input{conclusion.tex}

\bibliographystyle{elsarticle-num} 
\bibliography{references}





\end{document}

%% file: introduction.tex
\section{Introduction}
\label{sec:introduction}

To deliver a high-quality software product, it is essential to adopt systematic and quantifiable approaches across all software development lifecycle activities \citep{al2020agile}. Careful selection and implementation of appropriate software development approaches can significantly enhance the business value and user satisfaction of software systems \citep{rico2009business}. To ensure that the product's value and quality are maintained, it is imperative that software development activities are conducted in a formal and controlled manner. 

Software development is a complex process that involves a variety of approaches aimed at improving the quality and efficiency of software products. The literature and industry commonly employ various agile and traditional approaches, including Scrum, DevOps, Kanban, Extreme Programming (XP), Lean Development, Waterfall, and V-Model \citep{kuhrmann2017hybrid, kuhrmann2021makes}. These approaches are based on different concepts, such as plan-driven, iterative, incremental, or lean development \citep{kuhrmann2017hybrid}. Additionally, a hybrid development approach, defined as any combination of agile and traditional methodologies is utilized to design complex software systems \citep{kuhrmann2017hybrid, kuhrmann2021makes}. In the literature, various studies have focused on specific software development approaches. For example, Petersen et al. \citep{petersen2009waterfall} conducted an industrial evaluation of the waterfall model, Khan et al. \citep{khan2021agile} focused on agile trends in globally distributed software development environments, Kuhrmann et al. \citep{kuhrmann2019walking} examined software development approaches in academia, and Riaz et al. \citep{riaz2019implementation} studied the challenges and benefits of Kanban implementation.

However, to the best of our knowledge, there is a paucity of research exploring different aspects of software development approaches using data mining and analysis of relevant information obtained from Q\&A online platforms such as Stack Exchange (SE) \footnote{https://stackexchange.com/ accessed on 3/04/2022}, that includes SO \footnote{https://stackoverflow.com/ accessed on 3/04/2022}, SESE \footnote{https://softwareengineering.stackexchange.com/ accessed on 3/04/2022}, and PMSE \footnote{https://pm.stackexchange.com/ accessed on 3/04/2022}. These online platforms offer a valuable source of knowledge for development teams, where practitioners engage in discussions on diverse software development approaches, challenges, and concerns \citep{ali2020conceptualising, khan2019analysis}.

Recently, researchers have used supervised (deep \& machine learning) and unsupervised learning (topic Modelling) techniques to analyze a wide range of software development-related issues encountered by developers on Q\&A websites \citep{vidoni2022systematic}. These studies have proposed automated methods to extract developers' discussion posts on various topics, such as challenges in continuous software engineering \citep{zahedi2020mining}, identifying challenges in Docker development \citep{haque2020challenges}, security vulnerability \citep{le2021large}, software bug detection \citep{zhou2020improving}, developers' communications and their implications \citep{brisson2020we}, non-functional requirements \citep{paixao2017interplay}, design patterns \citep{dwivedi2018software}, and software maintenance and evolution \citep{sun2015msr4sm}. Nonetheless, we have not found any research that specifically investigates developers' discussions on Q\&A platforms with a focus on software development approaches. Therefore, this study aims to investigate and analyze the developers' discussion posts on core relevant SE platforms (including SO, SESE, and PMSE) in detail, aiming to provide valuable insights into software development approaches.

This study aims to fill this significant gap in the existing by critically analyzing and mining the topics, recent trends, and the challenges reported by the software developers across the three core platforms of SE previously mentioned. We assume that this work will help address common challenges developers face. It is harder for practitioners to decide where to focus their efforts without such information. Our research provides a significant contribution not only to practitioners but also to academic researchers. We have conducted an empirical study, collected 13903 software development approaches related to posts and used a mixed method approach that consists of Latent Dirichlet Allocation (LDA) \citep{blei2003latent} topic modeling and qualitative analysis to achieve the study objective. Furthermore, quantitive measurements are used to portray useful information for the software developers, which will help in improving software quality and user satisfaction by promptly incorporating the following research findings of this study:

\begin{enumerate}[(1)]
    \item The number of practitioners’ questions related to software development approaches has been on a gradual rise. However, the trend of developers’ questions (received accepted answers) in this research domain has been declining since 2014. 
    \item Fifteen commonly discussed software development approaches-related topics were identified using the LDA model. 
    \item Among the fifteen topics mentioned above, the popular and difficult ones were identified, and the top three most difficult topics exhibited an upward trend over time. A correlation analysis using Kendall's Tau correlation test \citep{kendall1938new} revealed a strong negative correlation between topic popularity and difficulty, indicating a decline in popularity as the difficulty of the topic increases. 
    \item We identified 49 challenges practitioners faced when deploying the software development approach, which were organized into 14 sub-themes and 4 higher-level themes. 
\end{enumerate}

This paper consists of the following seven sections. The study motivation is discussed in Section \ref{sec:motivation}. The related work to analyze the existing relevant studies is presented in Section \ref{sec:related-work}. Section \ref{sec:research-method} details the research methodology. The findings of this study are presented in Section \ref{sec:results} and the implications are reported in Section \ref{sec:implication}. The threats to the validity are highlighted in Section \ref{sec:threats}, and the conclusion, along with future avenues are presented in Section \ref{sec:conclusion}.

%% file: motivation.tex
\section{Motivation}
\label{sec:motivation}

Software companies, teams, and individual developers have been searching for effective and efficient software development approaches for decades. The Waterfall model, established in 1970, was the first mature development approach \citep{royce1987managing}, followed by the Spiral model \citep{boehm1988spiral}, and subsequently, Lean and Agile development approaches \citep{beck2001manifesto}, also including Scrum and present-day DevOps \cite{kim2021devops}. As software development activities rapidly evolve, an increasing number of development practices and tools have emerged, drawing from the concepts of the mentioned development approaches. To deliver high-quality products and stay competitive in the market, software practitioners must not only conduct development activities well but also possess knowledge of team management (e.g., the collaboration of developers and operations), time management (e.g., estimation of user stories), and tools usage (e.g., Kanban tools and Docker) \cite{sutherland2014scrum}. Selecting the appropriate development approach for an organization, team, or project is challenging, as software development approaches involve a complex set of practices \citep{kuhrmann20172nd}. Due to various development environments and complex software products, it is not feasible to have a single software development approach that fits all \citep{fraser2007no}. According to Klunder et al. \citep{klunder2017helena}, software practitioners do not always strictly follow defined development processes and practices. Their study highlighted that practitioners' perspectives on software development approaches differ from academic research and the unpredictable challenges that arise when using formal software development processes in an industrial setting. Marco et al. \citep{kuhrmann2017hybrid} suggested that practitioners use traditional development approaches as a framework and adopt agile methods at specific stages of the software development life cycle, resulting in hybrid development approaches \cite{kuhrmann2021makes}.

Therefore, understanding various aspects of software development approaches is crucial for software practitioners to tailor a development approach that aligns with their project's unique requirements. However, there is a lack of comprehensive knowledge and guidance that practitioners and researchers can use to achieve this understanding.  This gap prompted us to investigate the different aspects of software development approaches and provide a structured understanding that helps practitioners in enhancing software quality and contribute to the overall project success. In literature, various studies have investigated software development approaches and practices in the industrial domain \citep{kuhrmann2017hybrid, zhou2021system, aymerich2018software, mushashu2019investigating}. However, research specifically focusing on examining developers' discussions regarding software development approaches across Q\&A sites is lacking. Since software practitioners often seek information for resolving challenges from Q\&A sites, which contain a vast amount of data, exploring these sites with respect to software development approaches is crucial.

This research aims to investigate the discussion topics, trends and challenges of software development approaches from practitioners’ perspectives on the Stack Exchange (SE) platform. In particular, we aim to study:
\begin{itemize}
\item RQ1: How successfully the questions related to software development approaches are answered?
\item RQ2: What are the software development approaches discussion topics?  
\item RQ2.1: What are the most popular and difficult topics?
\item RQ3: What are the critical challenges practitioners encounter when implementing software development approaches? 
\end{itemize}

The insights gained from the study findings can aid researchers in advancing their investigations to improve software development activities. Furthermore, software practitioners can leverage the results to better understand the core aspects of software development approaches. 
Equipped with this understanding, developers can proactively prepare to establish advanced practices in software development and tackle challenges that may arise throughout the development process.

%% file: related-work.tex
\section{Related Work}
\label{sec:related-work}

We now summarized the related studies focusing on software development approaches (Section \ref{sec:related-work-software-development-approaches}), and  Q\&A software repositories mining (Section \ref{sec:related-work-mining}).

\subsection{Software development approaches}
\label{sec:related-work-software-development-approaches}
Software development approaches consistently evolve because of the rapid, iterative, and complex software development environments and domain problems. Bajec et al. \citep{bajec2007practice} addresses the issue of software development methods lacking adaptability to project-specific situations. They reveal that developers often avoid using methods that exist only on paper, as they fail to cater to the unique needs of different projects. The authors propose a new approach called "Process Configuration" to create project-specific methods from existing ones, considering the project's requirements. The proposed approach offers increased flexibility and is easier to implement compared to other existing approaches.

Marco et al. \cite{kuhrmann2021makes} conducted a large-scale international survey to investigate the factors that make a software development method agile. They analyze the perceived degree of agility in various project disciplines, development methods, and practices. Findings indicate that most projects exhibit increasing degrees of agility, with the selection of practices having a stronger impact than the methods used. The study concludes that agility cannot be defined solely at the process level, and additional factors must be considered when implementing or improving agility in a software company.

Bustard et al. \citep{bustard2013maturation} conducted an industrial survey to observe the principles and practices of agile development approaches adopted in 2010 and 2012. The research offers insights into the nature and practice of agile development, highlighting key outcomes and trends. The study findings further reveal that agile practices had been widely used before 2012. Since then, there has been a growing tendency to adopt agile methods. 

The more recent trend of integrating agile and traditional development approaches (i.e.hybrid development approaches) is becoming more common in practice. Tell et al. \citep{tell2021towards} explores the construction of hybrid software development methods by analyzing 1467 data points from a large-scale practitioner survey. The findings reveal that modern software development consists of eight core approaches and a few practices. The study proposes a systematic construction approach for hybrid methods, characterized by the practices they include. Using an 85\% agreement level, the researchers present examples of hybrid methods and introduce an initial construction procedure to define a method frame and incrementally enrich it with ranked sets of practices.

Additionally, the HELENA study conducted by Klunder et al. \citep{klunder2017helena} gained insights into the distribution of hybrid approaches. Preliminary findings suggest that combining traditional and agile software development approaches provides an opportunity to deliver a software product in a continuous loop, get frequent feedback, and improve overall productivity. This study's findings only cover the German industrial domain.

Khan et al. \cite{khan2021agile} conducted an industrial study aims to develop a taxonomy of factors that positively impact the scaling process of agile methods in the Chinese Global Software Development (GSD) industry. Factors are identified through a literature review and an industrial empirical study with Chinese agile and GSD practitioners. The resulting factors are categorized, prioritized, and organized into a taxonomy using the Fuzzy AHP multi-criterion decision-making approach. This taxonomy provides a valuable resource for the GSD industry to assess and improve the scaling process of agile methods.

\subsection{Mining software issues across Q\&A repositories}
\label{sec:related-work-mining}
Various studies used an unsupervised learning approach (topic modeling) to mine software development-related issues from different repositories. 

For instance, Barua et al. \citep{barua2014developers} used the LDA approach to analyze the textual content of SO discussions and automatically discover the main topics present in developer conversations. This approach differs from prior work, which focused on user activities or social interactions in Q\&A websites. By analyzing the discovered topics, their relationships, and trends over time, the authors gained valuable insights into the development community. They observed a wide range of topics of interest to developers, including jobs, version control systems, and C\# syntax. Additionally, they found that some questions led to discussions on other topics and that the most popular topics over time were web development (particularly jQuery), mobile applications (especially Android), Git, and MySQL. This analysis helps the software engineering community better understand the thoughts and needs of developers.

In a continuous software engineering (CSE) domain, Zahedi et al. \citep{zahedi2020mining} used the LDA approach to empirically investigate the practitioners' perspectives by mining Q\&A discussions on the SO platform. They used a topic modelling approach to identify dominant topics and conduct qualitative analysis to pinpoint key challenges. The study found that questions are becoming more technology-specific and harder to answer. Among the 32 identified topics, "Error messages in Continuous Integration/Deployment" and "Continuous Integration concepts" were the most dominant. The paper also highlights the most challenging areas in CSE from practitioners' viewpoints.

Haque et al. \citep{haque2020challenges} conducted a 
a large-scale empirical investigation on Docker technology by mining 113,922 Docker-related posts from the SO community. Using the LDA approach for topic modeling, the authors identified 30 topics grouped into 13 main categories, with the majority of posts belonging to application development, configuration, and networking categories. The study found that monitoring status, transferring data, and authenticating users were particularly popular topics among developers. It also revealed challenges faced by developers in areas such as web browser issues, networking errors, and memory management, as well as a lack of experts in the domain. The findings are expected to guide future research on the development of new tools and techniques and help the community focus their efforts on Docker-related topics.

Minh Le et al. \citep{le2021large}  analyse developers' Security Vulnerabilities (SVs) discussions on two major Q\&A websites, SO and Security StackExchange (SSE). Using topic modeling, they examined 71,329 SV posts to identify 13 main discussion topics. The study found that these topics did not necessarily align with expert-based security sources like Common Weakness Enumeration (CWE) and Open Web Application Security Project (OWASP). The analysis also revealed that while SV discussions tend to attract more expert answers than other domains, some difficult topics, such as Vulnerability Scanning Tools, receive limited expert support. Furthermore, the authors identified seven key types of answers to SV questions, with SO often providing code and instructions, while SSE offers experience-based advice and explanations. The findings aim to help researchers and practitioners effectively acquire, share, and leverage SV knowledge on Q\&A websites.


Vidoni \citep{vidoni2022systematic} identified the need for guidelines on conducting systematic Mining Software Repositories (MSR) studies, as existing research often lacks a systematic approach for repository selection and data extraction. They conducted a systematic literature review of MSR studies. The results showed that many MSR studies do not report selection or data extraction protocols and rarely discuss threats to validity due to the selection or data extraction steps. The authors concluded that there is a need for guidelines on conducting systematic MSR studies and proposed new guidelines and a template to consolidate related studies and strategies for systematic literature reviews in the MSR field.

 
 \subsection{Comparative analysis}
 
 Our study aims to explore software development approaches across Q\&A forums. We discussed several related studies that investigate different aspects of software development, none of them specifically focus on the exploration of Q\&A forumes (repositories) for software development approaches. Previous research has analyzed different software development approaches, such as agile, traditional, and hybrid methods, and their impacts on project success, agility, and productivity. Additionally, several studies have employed unsupervised learning techniques, such as topic modeling (see Section \ref{sec:research-method}), to mine software development-related issues from Q\&A repositories. However, these studies focused on specific technologies or challenges within the software development domain rather than the development approaches themselves.

Our study's significance lies in filling the gap in the existing literature by investigating software development approaches through the lens of Q\&A websites. By doing so, we aim to gain valuable insights into developers' thoughts, preferences, and challenges related to different software development approaches, which could help inform and improve future development practices.

%% file: research-method.tex
\section{Research Method}
\label{sec:research-method}

\subsection{Research questions}
\label{sec:research-method-questions}
This study investigated four research questions to identify common software development approaches topics, trends, and challenges to help software practitioners improve their development activities. To answer these RQs, we retrieved 13903 software development approaches related to posts from SO, SESE, and SEPM using the approach described in Section \ref{sec:research-method-data-collection}.

\textbf{RQ1: How successfully the questions related to software development approaches are answered?}

\emph{\underline{Rationale:}} Our study's first research question (RQ1) examines the nature of responses to questions related to software development approaches. The findings from RQ1 can provide insights to identify areas where more clarification or information is needed, which can lead to more targeted investigations and a better understanding of the subject matter.

\emph{\underline{Method:}} Initially, we computed the average number of answers for questions related to software development approaches and compared it to the respective value, taking inspiration from Zahedi et al. \citep{zahedi2020mining}. Additionally, we analyzed the question that received the most answers to gain a more comprehensive understanding of the success rate of addressing questions in the software development approaches domain.

Drawing on previous research \citep{zahedi2020mining, pinto2014mining, treude2011programmers}, we classified questions related to software development approaches into three categories:
\begin{itemize}
\item Successful questions: those that received an accepted answer
\item Ordinary questions: those that received answers but no accepted answer
\item Unsuccessful questions: those that received no answers
\end{itemize}
We then calculated the distribution of these categories and illustrated the growth trends for all questions, successful questions, ordinary questions, and unsuccessful questions within this domain.

\textbf{RQ2: What are the software development approaches discussion topics?}

\emph{\underline{Rationale:}} Identifying the key discussion topics across Q\&A repositories provides a comprehensive understanding of the software development landscape, allowing practitioners and researchers to stay up-to-date with current trends and practices. Additionally, by understanding the various discussion topics, practitioners can make well-informed decisions regarding the selection and implementation of suitable software development approaches based on project requirements and constraints.

\emph{\underline{Method:}} Building upon the work of Mansooreh et al. \citep{zahedi2020mining}, we employed the Latent Dirichlet Allocation (LDA) \cite{blei2003latent} topic modeling technique  (Section \ref{sec:research-method-lda}) to identify most common discussion topics related to software development approaches. Topic modeling is a Natural Language Processing (NLP) technique that automatically extracts structured information from a collection of documents \citep{chen2016survey}. For the LDA model, we considered a question's title, body, and corresponding answers as a single input document, and the output consisted of the most frequently occurring topics identified in the text corpus.

\textbf{RQ2.1: What are the most popular and difficult topics?}

\emph{\underline{Rationale:}}  Identifying popular and difficult topics enables practitioners and researchers to focus their efforts on areas that require the most attention, leading to better outcomes in software development projects. Moreover, difficult topics may uncover opportunities for innovation, as researchers and practitioners can develop new techniques, tools, or processes to address challenges and streamline software development processes.

\emph{\underline{Method:}} Drawing from the work of \citep{ahmed2018concurrency, rosen2016mobile, le2021large}, we utilized average values of (P1) views, (P2) score, (P3) favorite count, and (P4) comments to gauge the popularity of developers' topics. Concurrently, we computed three metrics—(D1) percentage of accepted answers, (D2) median duration (minutes) to receive an accepted answer since creation, and (D3) average percentage of answers to views—to assess topic difficulty. Intuitively, a more popular developers' topic would exhibit higher average views, scores, favorite count, and comments, and vice versa. In contrast, a more challenging topic would have a higher D2 value but lower D1 and D3 difficulty metric values.

To generate a more uniform and representative value across various developers' topics, we calculated the reciprocals of D1 and D3 and then determined the geometric mean (Equation 6) of popularity (Equation 1) and difficulty (Equation 2) metrics. We opted for the geometric mean over the arithmetic mean due to the different metric units. To enhance the comparison of all topics' popularity and difficulty, we normalized the popularity (Equation 4) and difficulty values (Equation 5) using Min-max normalization (Equation 5).
\begin{equation}
    Popularity_i=G(P1_i,P2_i,P3_i,P4_i ) \label{con:popularity} 
\end{equation}
\begin{equation}
    Difficulty_i=G(1/D1_i ,D2_i,1/D3_i ) \label{con:difficulty} 
\end{equation}
\begin{equation}
    Normalized\_Popularity_i=N(Popularity_i ) \label{con:normalized-popularity} 
\end{equation}
\begin{equation}
    Normalized\_Difficulty_i=N(Difficulty_i ) \label{con:normalized-difficulty} 
\end{equation}
\begin{equation}
    N(x_i) = \frac{x_i - min(x)}{max(x) - min(x)} \label{con:normalization} 
\end{equation}
\begin{equation}
    G(x_1,x_2,\cdots,x_n) = (\prod_{i=1}^{n}x_i)^\frac{1}{n} \label{con:geometric} 
\end{equation}

\textbf{RQ3: What are the critical challenges practitioners encounter when implementing software development approaches?}

\emph{\underline{Rationale:}} RQ3 identified the critical challenges faced by practitioners concerning various activities within software development approaches. By being aware of the challenges, practitioners can make informed choices when selecting and adopting software development approaches, considering potential obstacles, and planning for contingencies to ensure project success.

\begin{figure*}[htbp]
  \centering
  \includegraphics[width=\linewidth]{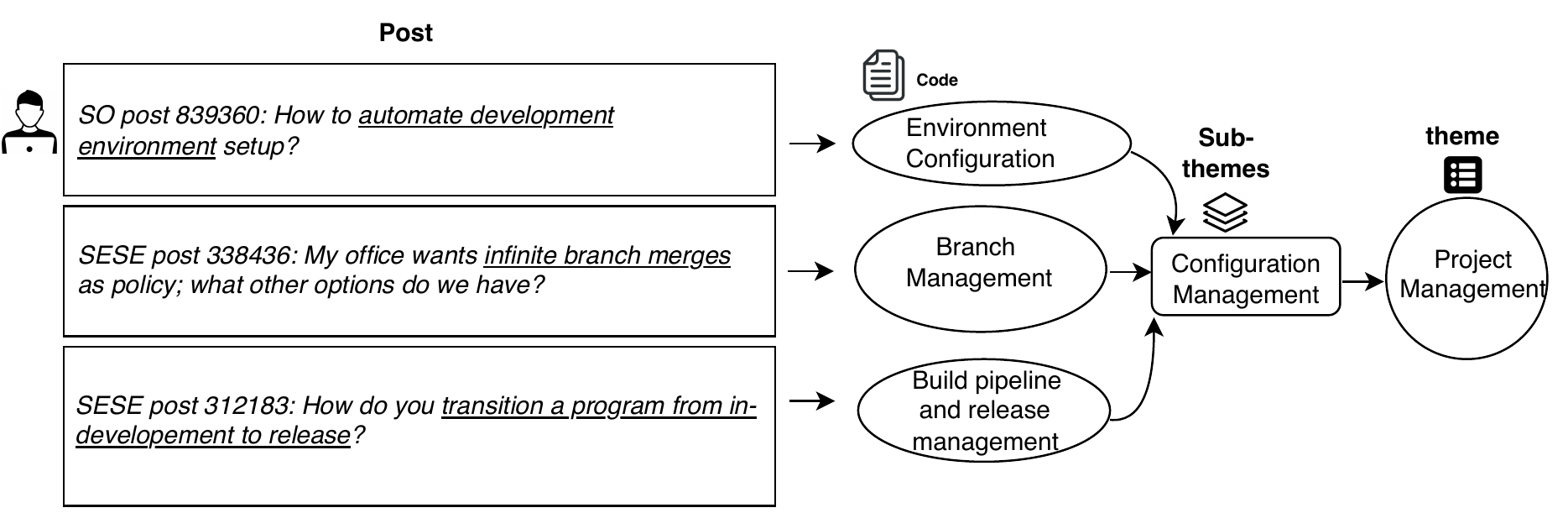}
  \caption{Translation of codes into themes}
  \label{Fig.theme-sample}
\end{figure*}

\emph{\underline{Method:}} We firstly adopted the Accumulated Post Score (AMS) formula proposed by Bajaj et al. \citep{bajaj2014mining} to rank the collected developer’s posts:
$$AMS_i=3U_i-25D_i+10C_i+A_i+F_i$$
Where $U_i$, $D_i$, $C_i$, $A_i$, and $F_i$ represent $developers Post_i$ upvotes, downvotes, comment count, answer account, and favorite count, respectively. The equation considers the above factors to rank the developer's posts and identify their importance. By computing AMS for all collected developer posts in the data set, we selected the top 200 questions with the highest score for manually analyzing and evaluating. We used the thematic analysis approach proposed by Braun and Clarke \citep{braun2006using} for qualitative data analysis. The thematic synthesis in software engineering research is considered flexible and can be used to produce an insightful synthesis \citep{cruzes2011recommended}. The first and second authors manually open-coded the documents using the MAXQDA \footnote{https://www.maxqda.com/} tool. The third and fourth authors cross-checked and updated the codes based on their understanding. Finally, any differences in opinion were discussed and agreed upon in a joint meeting by inviting all the authors. We identified 49 initial codes, and then these emerging codes were further mapped into 14 sub-themes and 4 higher-order themes. The sample analysis of the proposed thematic synthesis process is depicted in Figure \ref{Fig.theme-sample}.

\subsection{Data collection}
\label{sec:research-method-data-collection}
To examine the support provided by Q\&A sites for discussions on software development approaches, we gathered research data from three popular software Q\&A repositories: SO, SESE, and SEPM. Stack Overflow (SO) is the best-known online platform for developers to ask questions, learn, and share their programming knowledge. Stack Exchange Software Engineering (SESE) is an online community for academics and developers to ask questions related to the software development activities. Additionally, Stack Exchange Project Management (SEPM) is a public platform for project managers to ask questions and receive answers. Many developers post software development-related queries on these platforms to obtain recommendations and frequent assistance from experts. Such information can help software practitioners improve the existing software development process and enhance product quality.

For this purpose, we collected commonly used software development approaches names (e.g., Scrum, Domain-Driven Design, and DevOps) from \cite{zhou2021system, klunder2017helena, kuhrmann2017hybrid, kuhrmann2019walking} and other custom pivotal related concepts (e.g., development process, agile, and software development life cycle) as potential tag keywords. Based on these keywords, we identified 28 valid tags from SO, SESE, and SEPM. The details of the data collected from SO, SESE, and SEPM regarding developers' questions related to the software development approaches are shown in Table \ref{tab:tags}. The column “selected tags" represents the software tags against which we collected developers' questions from SO, SESE, and SEPM. The columns “SO", “SESE" and “SEPM questions" display the number of developer queries collected against each tag. The column “Total" indicates the total number of developer questions across the three Q\&A platforms. We excluded developers' posts containing code snippets in their bodies to prevent data from leaning towards a specific technology and code language directions. Furthermore, we removed repeated posts with multiple selected tags. In summary, we extracted 13,903 developers' questions along with their corresponding answers using the software tags mentioned in Table \ref{tab:tags} from the Stack Exchange Data Dump \footnote{https://archive.org/details/stackexchange\_20211206} downloaded from the “Internet Archive." The data was retrieved on July 6, 2022.

    \begin{table}[htbp]
        \footnotesize
        \centering
      \caption{Selected Tags and Question Numbers (exclude questions that contain code snippet)}
      \label{tab:tags}
      \begin{tabular}{| l | l | p{1.6cm} | p{1.6cm} | p{1.6cm} | p{1.6cm} |}
        \hline
        \textbf{No.} & \textbf{Selected Tags} & \textbf{SO questions} & \textbf{SESE questions} & \textbf{PMSE questions} & \textbf{Total} \\
        \hline
        1 & domain-driven-design & 3507 & 835 & N/A & 4497 \\
        2 &		agile &	965 &	1093 &	1351 &	3410 \\
        3 &		devops &	3156 &	73 &	21 &	3342 \\
        4 &		Scrum &	729 &	717 &	1534 &	2981 \\
        5 &		development-process &	31 &	676 &	104 &	813 \\
        6 &		kanban &	173 &	47 &	309 &	532 \\
        7 &		waterfall &	72 &	46 &	68 &	191 \\
        8 &		sdlc &	64 &	88 &	11 &	163 \\
        9 &		extreme-programming &	45 &	55 &	17 &	117 \\
        10 &		agile-project-management &	77 &	N/A &	N/A &	77 \\
        11 &	 	prince2 &	N/A &	3 &	56 &	59 \\
        12 &		agile-processes &	53 &	N/A &	N/A &	53 \\
        13 &		mda &	47 &	N/A &	N/A &	47 \\
        14 &		kanban-board &	N/A &	N/A &	42 &	42 \\
        15 &		model-driven-development &	39 &	N/A &	N/A &	39 \\
        16 &		rational-unified-process &	18 &	13 &	N/A &	31 \\
        17 &		rup &	23 &	N/A &	5 &	28 \\
        18 &		safe &	N/A &	N/A &	24 &	24 \\
        19 &		iterative-development &	N/A &	22 &	N/A &	22 \\
        20 &		Lean &	N/A &	22 &	N/A &	22 \\
        21 &		scrumban &	N/A &	N/A &	17 &	17 \\
        22 &		dsdm &	N/A &	N/A &	12 &	12 \\
        23 &		scaled-agile-framework &	N/A &	9 &	N/A &	9 \\
        24 &		personal-software-process &	7 &	2 &	N/A &	9 \\
        25 &		nexus &	N/A &	N/A &	7 &	7 \\
        26 &		feature-driven &	3 &	N/A &	N/A &	3 \\
        27 &		large-scale-scrum &	N/A &	3 &	N/A &	3 \\
        28 &		dsdm-atern &	2 &	N/A &	N/A &	2 \\
        \hline \hline
        & Total (Remove Duplicates) & 8486 &	2978 &	2439 &	13903 \\
        \hline
      \end{tabular}
    \end{table}

\subsection{Topic Modelling}
\label{sec:research-method-lda}
Topic Modeling is a Natural Language Processing (NLP) technique that automatically extracts structured information, such as topics, from a corpus of documents \citep{chen2016survey}. A set of semantically related words that frequently co-occur in textual data can be considered a representation of one topic \citep{chen2016survey, blei2003latent}. Using a topic modeling framework to analyze a corpus of documents allows researchers to efficiently organize and index documents based on their semantic structure \citep{chen2016survey}. Latent Dirichlet Allocation (LDA) \citep{blei2003latent} is among the most popular unsupervised topic modeling algorithms frequently employed to extract topics from text corpus \citep{barua2014developers}. It has been used in different other domains \citep{griffiths2004finding, jacobi2016quantitative}, such as continuous software engineering \citep{zahedi2020mining}, Docker development \citep{haque2020challenges}, requirements engineering \citep{khan2018linguistic}, and security vulnerability \citep{le2021large}. Following the work of Mansooreh et al. \citep{zahedi2020mining}, we define a question's title, body, and corresponding answers as one input document for the LDA model and output the number of frequently occurring topics identified in the text corpus.

\subsubsection{Preprocessing of collected posts}
Before applying LDA to the data set curated from the developer's queries on the SO and SE platforms, we need to preprocess the collected textual documents to remove the noise. For this purpose, we perform the following steps:
\begin{itemize}
    \item First, we removed HTML tags, code snippets, punctuation, and stop words using the NLTK \footnote{https://github.com/nltk/nltk} stopwords corpus.
    \item Then we converted all document texts to lowercase.
    \item We subsequently performed lemmatization (keeping only noun, adj, vb, and adv) and word stemming (converting words to their root form) to remove multi-form and irrelevant words.
    \item Finally, we adopted the approach proposed by Yang et al. \citep{yang2016security} that made the word frequency statistics of the data corpus and excluded the words which occurred less than 10 times.
\end{itemize}

\subsubsection{Topic modelling with LDA}
After preprocessing the dataset, we applied the LDA algorithm to the collected textual documents of developers' discussions on the Q\&A platforms. The number of topics "K" extracted from the training corpus is one of the most important input parameters in implementing LDA; if K is too small, the recovered topics may overlap, which is difficult to generalize. On the contrary, LDA may create excessively fine-grained topics in case of K is too large. In this regard, we observed a consecutive range of K from 2 to 50 with an increment of 1. Similar to \citep{barua2014developers, ahmed2018concurrency, rosen2016mobile, le2021large}, alongside K, we also set hyper-parameters ($\alpha$ and  $\beta$) values with an inclusive range from 0.01 to 1 in the step of 0.2, and an additional value “symmetric” ($1.0/numTopics$) and “asymmetric” ($1.0/(topicIndex+\sqrt{numTopics})$) for $\alpha$, “symmetric” for $\beta$. $\alpha$ determines the sparsity of document-topic distribution, and $\beta$ controls the sparsity of topic-word distribution. As proposed in other research studies \citep{rosen2016mobile, abdellatif2020challenges}, we measured the coherence score to choose the optimal number of identified topics because it highly correlates with human comprehensibility \citep{roder2015exploring}. Topic coherence represents the semantic correlation between high-scoring words that appeared in the topic. Hence, we trained the LDA model with each tuple of (K, $\alpha$, $\beta$) and chose the top 5 highest coherence values (with different K values) with corresponding results for comparison. To validate the results, the first author manually checked the topic words and most related developers' posts for 5 candidate K values to assure the optimal K value was selected. After these steps, we found that a tuple value of (15, 0.41, 0.81) provides relatively granular topics for collected software development processes-related developers discussion in the Q\&A platforms (SO and SE). Significantly, we filtered out topics with a probability less than 0.1 in a document to exclude unimportant topics adopted from \citep{barua2014developers}. To give each discussion topic descriptive labels in the Q\&A platforms, we carefully inspected the top 10 most frequent terms. We read through the top 15 developer discussion posts with the highest relevance to each topic, as reported in \citep{ahmed2018concurrency, le2021large, zahedi2020mining}. We also provide the complete list of topics, top 10 words, top 15 posts, and assigned labels \citep{peng2022topics}.

%% file: results.tex
\section{Results}
\label{sec:results}

In this section, we summarize the results of the data analysis, which aims to address the core research questions of this study comprehansively.

\begin{figure*}[htbp]
  \centering
  \includegraphics[width=\linewidth]{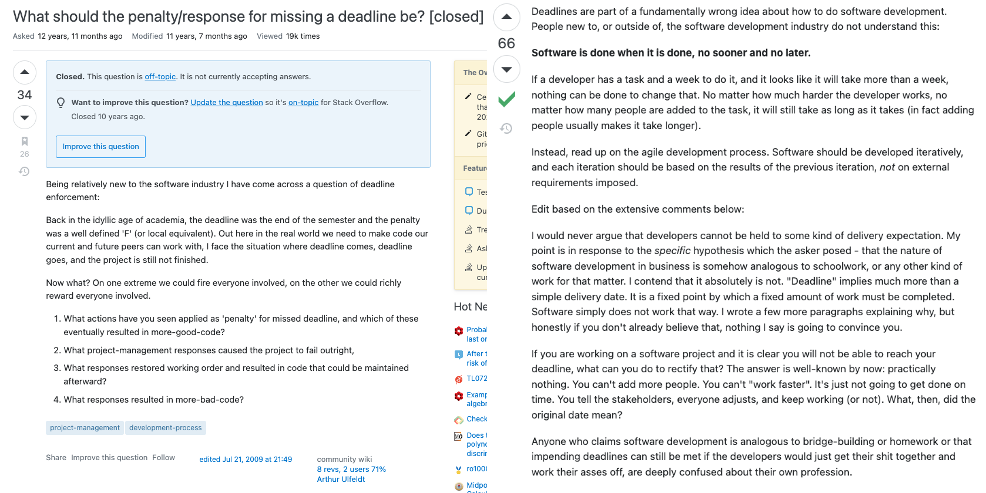}
  \caption{The question that received maximum answers}
  \label{Fig.sample-post}
\end{figure*}

\subsection{RQ1: Successfully answered questions}

We investigated how frequently the developers responded to questions related to software development approaches by mining their answers from Q\&A platforms (SO, SESE, and SEPM). Upon performing statistical analysis of practitioners questions that received answers, we observed that most questions in the software development approaches domain received fewer than three answers (mean=2.41). However, several developers' questions attracted considerable attention from practitioners, with a high number of responses. The developer question and its accepted answer which received a maximum of 37 responses from practitioners are presented in Figure \ref{Fig.sample-post}. We can observe that this question was published in 2009 on SO and pertained to project management regarding deadlines in software projects. The accepted answer to this question received many upvotes and stated,  “Software is done when it is done, no sooner and no later." Interestingly, this question was closed in 2012 as it did not fit the specific topics on SO.

\begin{table}
\centering
\caption{Distribution of Questions (Successful, Ordinary, Unsuccessful)}
\label{tab:distribution}
\small 
\begin{tabular}{|l|l|l|l|l|}
\hline
        & Successful & Ordinary & Unsuccessful & Total \\ \hline
\#Posts & 7299       & 5733     & 871          & 13903 \\ 
\%Posts & 52.50\%    & 41.24\%  & 6.26\%       & 100\% \\ \hline
\end{tabular}
\end{table}

\begin{figure*}[h]
  \centering
  \includegraphics[width=0.8\linewidth]{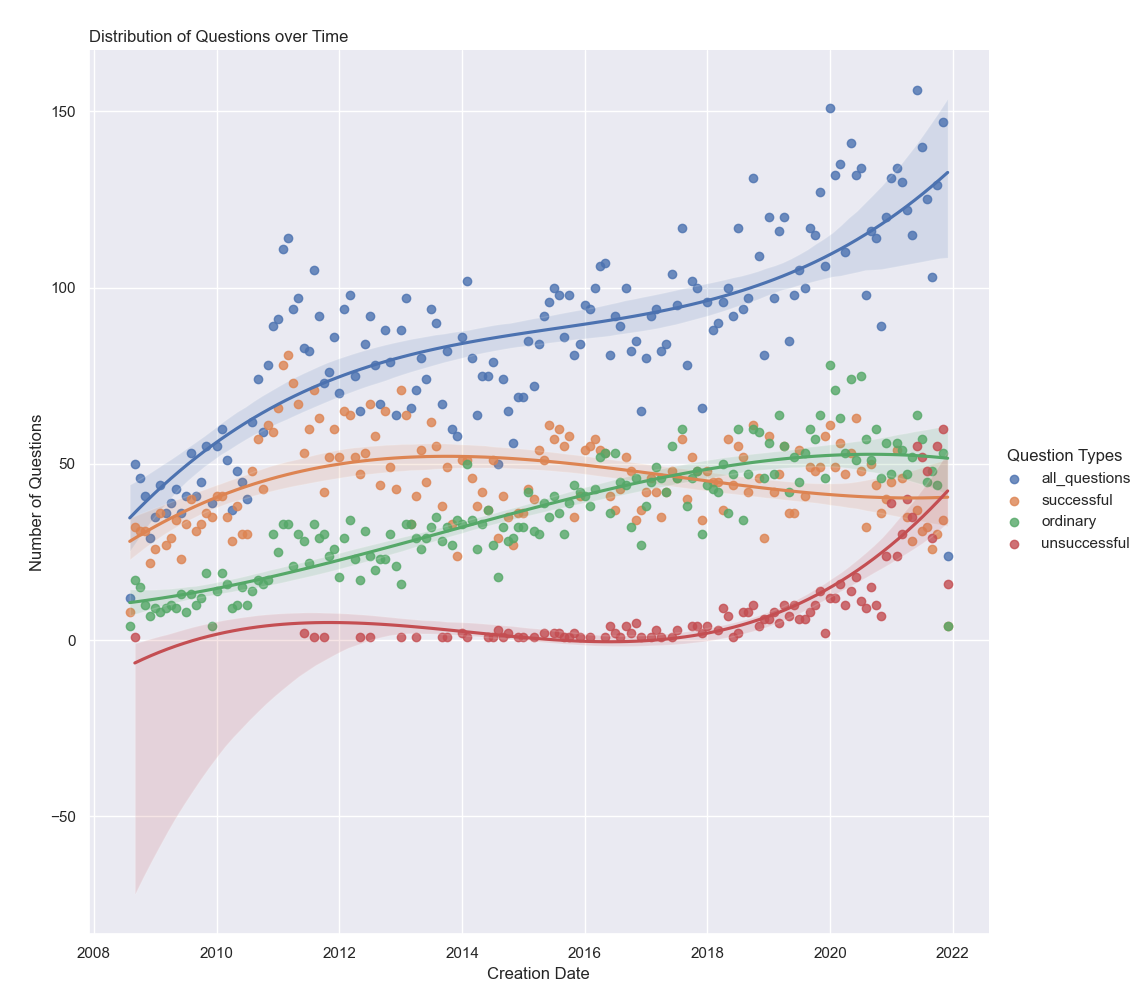}
  \caption{Trend of Questions}
  \label{Fig.successful}
\end{figure*}


We categorized software development approach-related questions into three types: successful, ordinary, and unsuccessful questions, as outlined in Section \ref{sec:research-method-questions} (RQ1). The distribution of these categories can be found in Table \ref{tab:distribution}. Interestingly, we can conclude that most of the questions, i.e., 7299 (52.50\%) in the software development approaches domain are identified as successful. Also, only 871 (6.26\%) developers' questions are categorized as unsuccessful and have not received any answer from the community on the SO and SE platforms. In contrast, 5733 (41.24\%) developers' posts are classified as ordinary  questions, where they received responses from developers but not a successful answer (See Table \ref{tab:distribution}). Furthermore, we depicted the trends of the developers' questions (successful, ordinary, and unsuccessful) in Figure \ref{Fig.successful}.

It can be concluded that the number of questions in the software development approaches domain has maintained sustainable and stable growth. This observation indicates that software development approaches continuously evolve from plan-based (traditional) to agile development, and still, guide practitioners toward new directions (e.g., hybrid, DevOps). From 2014, the trend of successful questions continued to decline and remained under ordinary questions since 2017. This phenomenon may suggest that the software development approaches domain needs more attention from practitioners and senior experts on open-source Q\&A platforms. There has been a drastic increase in the trend of unsuccessful questions since 2018. This trend might signify that the questions in this domain have become more complex, advanced, and challenging to answer.

\begin{tcolorbox}[colback=gray!5!white,colframe=gray!75!black,title=Key Findings of RQ1]
\textbf{Finding 1}: In the software development approaches domain, 52.50\% of questions were successfully answered on Q\&A platforms. However, 6.26\% were deemed unsuccessful as they received no responses from the community.

\textbf{Finding 2}: The success rate of software development approach questions has declined since 2014, indicating a need for more expert input on Q\&A platforms and a potential rise in question complexity.
\end{tcolorbox} 

\subsection{RQ2: Software development approaches topics}\label{RQ2: Software Development Approaches Topics}
\begin{figure*}[htbp]
  \centering
  \includegraphics[width=\linewidth]{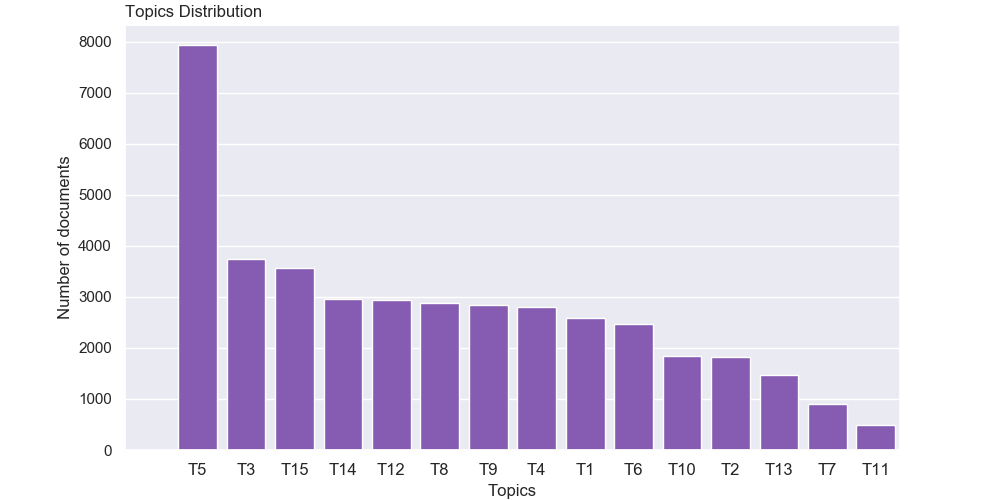}
  \caption{Distribution of Dominant Topics}
  \label{Fig.distribution}
\end{figure*}

To answer RQ2, we used topic modeling LDA on the collected textual documents of developers' posts from SO, SESE, and SEPM. Following the steps described in Section \ref{sec:research-method-lda}, we identified the optimal topics number (K = 15) and parameters ($\alpha$ = 0.41, $\beta$ = 0.81) by comparing the coherence score, manually checking the topics keywords, and corresponding documents. After training the LDA model, we analyzed and inspected the top 10 keywords identified by the LDA algorithm and read through the top 15 developers' posts with the highest relevance to each topic. The relevant developers topics were manually selected based on the identified keywords using the LDA algorithm. After performing the above steps, we gave each developer topic a descriptive label, as shown in Table \ref{tab:topics-trends}. The complete list of developers topics, top 10 words identified using the LDA algorithm, top 15 practitioners posts’ Id and assigned labels could be found at \citep{peng2022topics}. To filter the unimportant topics, we considered a valid topic in a document whose probability related to the document was at least 10\% \citep{barua2014developers}. Table \ref{tab:topics-trends} presents a comprehensive list of the identified topics, while Figure \ref{Fig.distribution} illustrates the frequency of occurrence of each topic. Our analysis revealed that “T5: Project, team and Time Management”, “T3: Domain model, design patterns and layers in DDD” and “T15: Software design and requirement” are respectively the top 3 highly asked topics in software development approaches related posts.

Each identified topic with sample developer's posts is elaborated as follows-  aimed at facilitating comprehension of their significance.

\begin{table}[htbp]
\centering
\caption{Software development approaches topics on SE identified by LDA along with their proportions and trends over time.}
\label{tab:topics-trends}
\resizebox{\textwidth}{!}{%
\begin{tabular}{|c|c|c|c|}
\hline
Topic & Topic Name                                               & \begin{tabular}[c]{@{}c@{}}Number (Proportion) \\ of questions\end{tabular} & Trend \\ \hline
T1  & Continues integration, build and deployment       & 2587 (18.61\%) & \begin{minipage}[b]{0.115\columnwidth}
		\centering
		\raisebox{-.5\height}{\includegraphics[width=\linewidth]{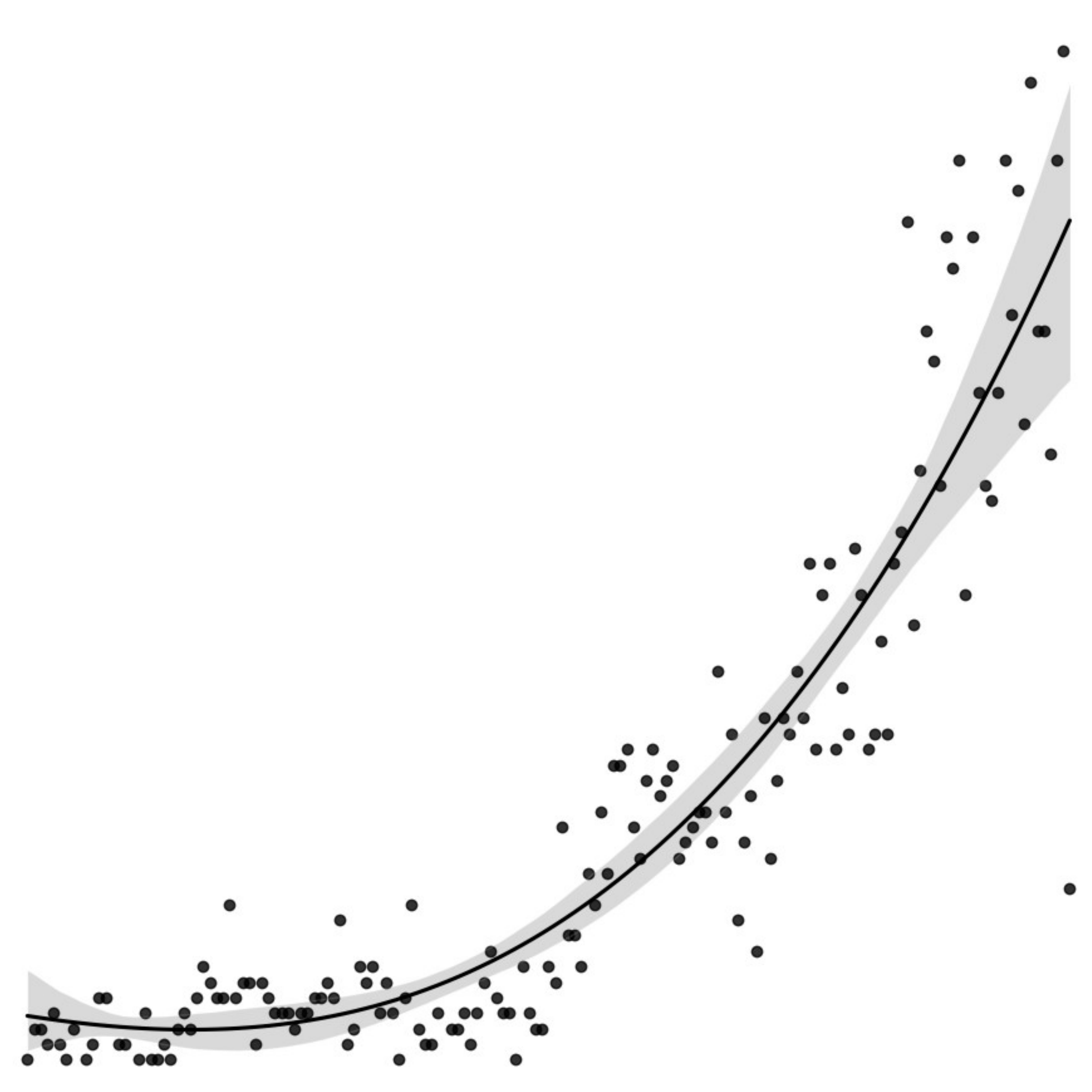}}
	\end{minipage} \\ \hline
T2  & Events, bounded contexts and Microservices in DDD & 1819 (13.08\%) &  \begin{minipage}[b]{0.115\columnwidth}
		\centering
		\raisebox{-.5\height}{\includegraphics[width=\linewidth]{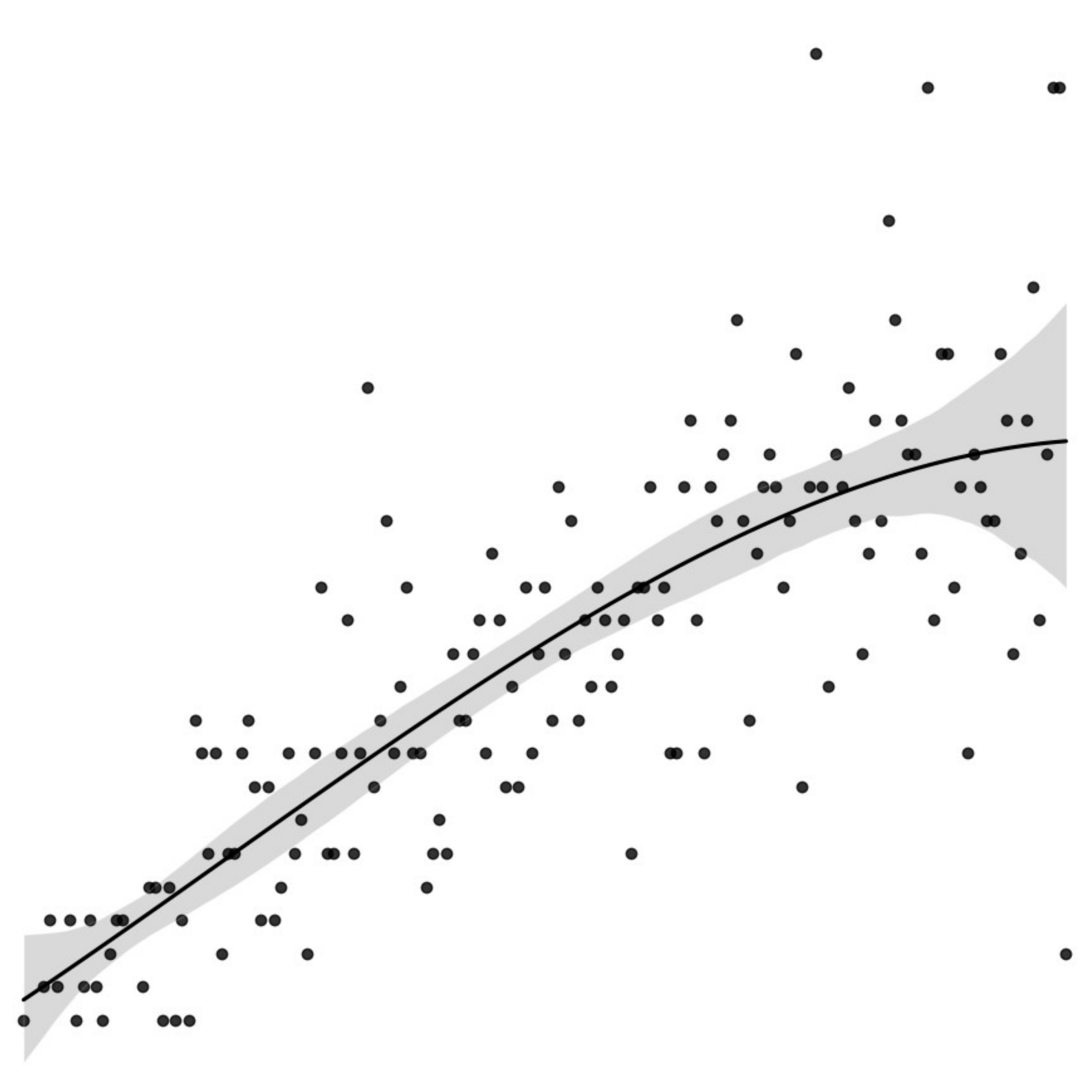}}
	\end{minipage} \\ \hline
T3  & Domain model, design patterns and layers in DDD   & 3740 (26.90\%) &  \begin{minipage}[b]{0.115\columnwidth}
		\centering
		\raisebox{-.5\height}{\includegraphics[width=\linewidth]{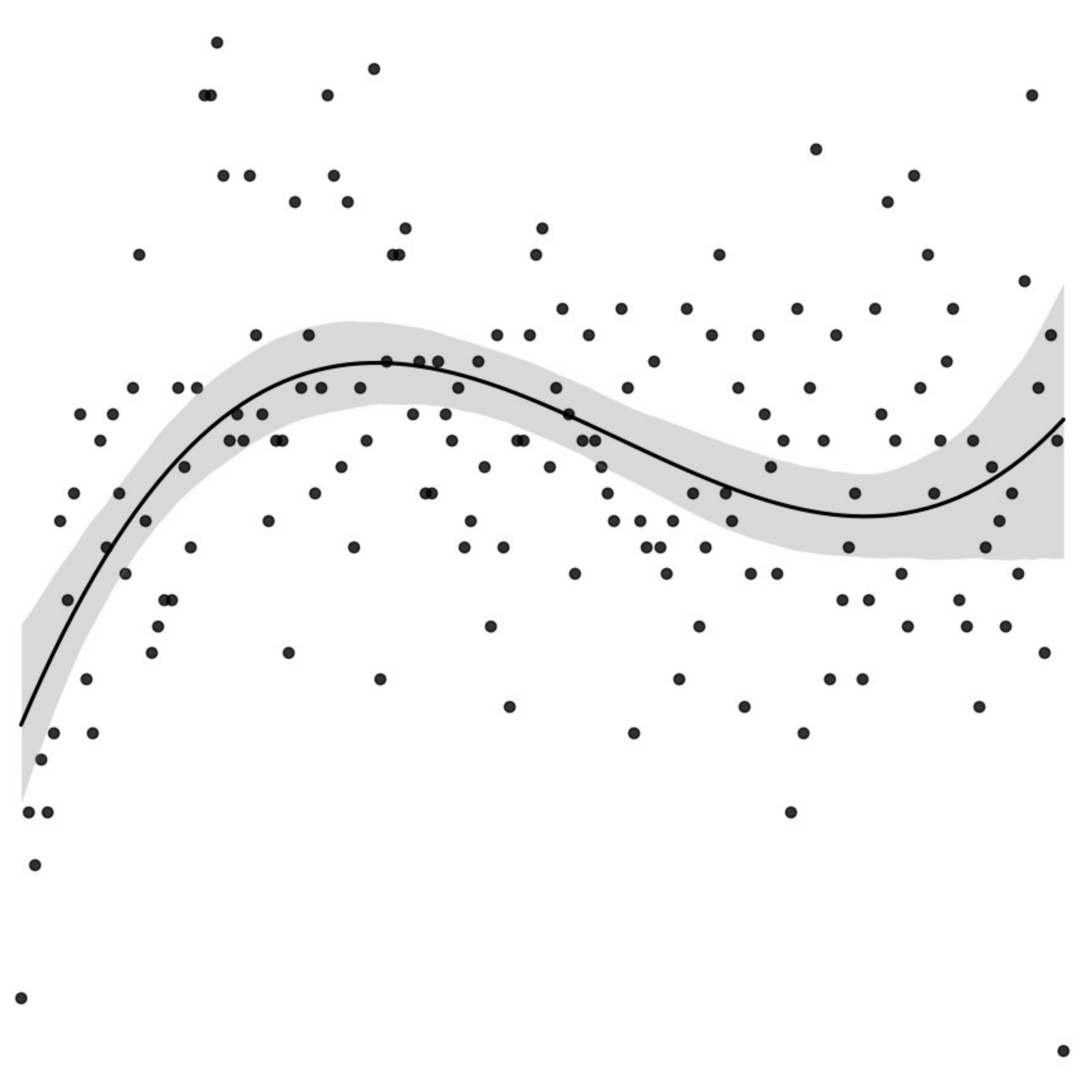}}
	\end{minipage} \\ \hline
T4  & DevOps automation tools                           & 2809 (20.20\%) &  \begin{minipage}[b]{0.115\columnwidth}
		\centering
		\raisebox{-.5\height}{\includegraphics[width=\linewidth]{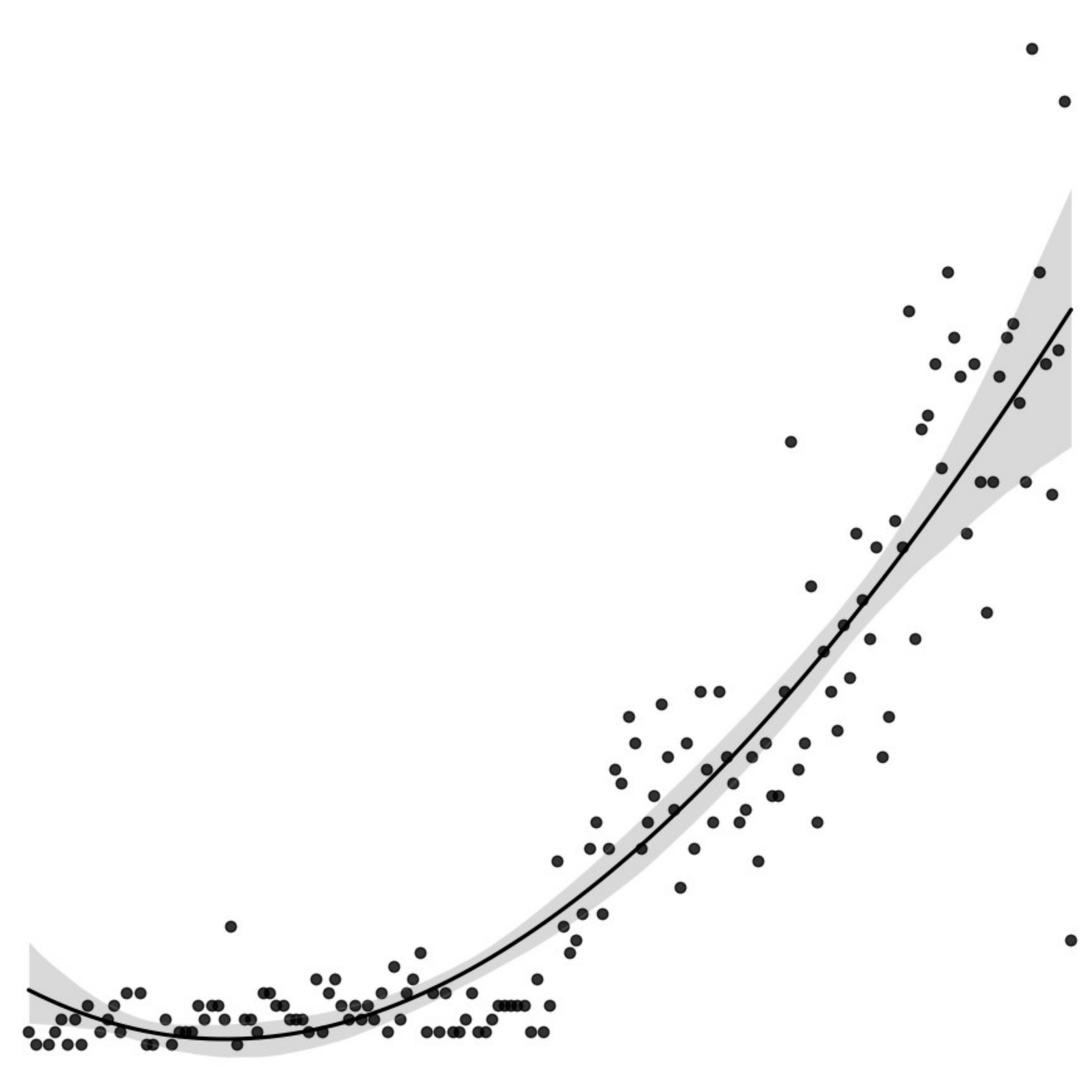}}
	\end{minipage} \\ \hline
T5  & Project, team and time management                 & 7933 (57.06\%) &  \begin{minipage}[b]{0.115\columnwidth}
		\centering
		\raisebox{-.5\height}{\includegraphics[width=\linewidth]{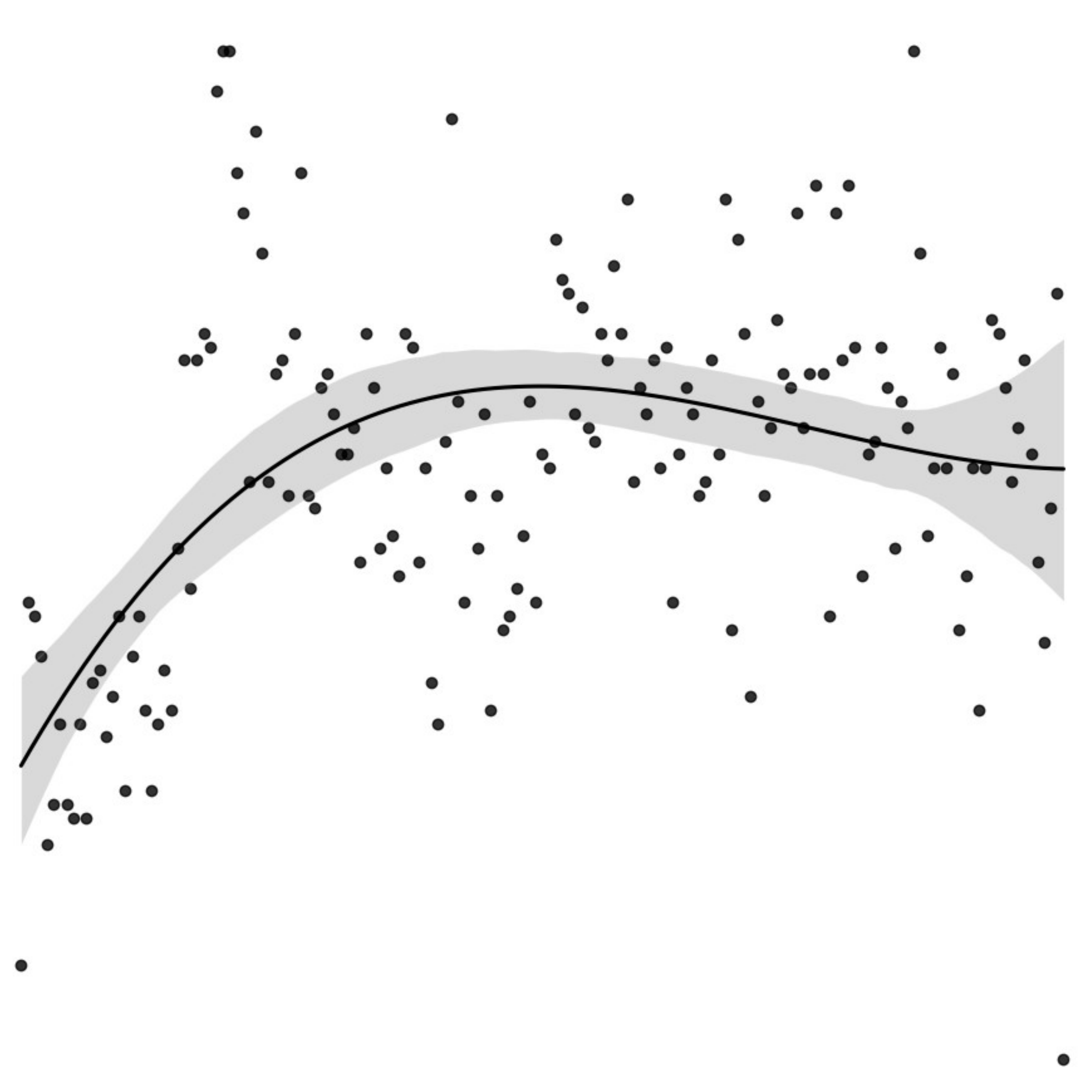}}
	\end{minipage} \\ \hline
T6  & Team role and responsibilities in Scrum           & 2462 (17.71\%) &  \begin{minipage}[b]{0.115\columnwidth}
		\centering
		\raisebox{-.5\height}{\includegraphics[width=\linewidth]{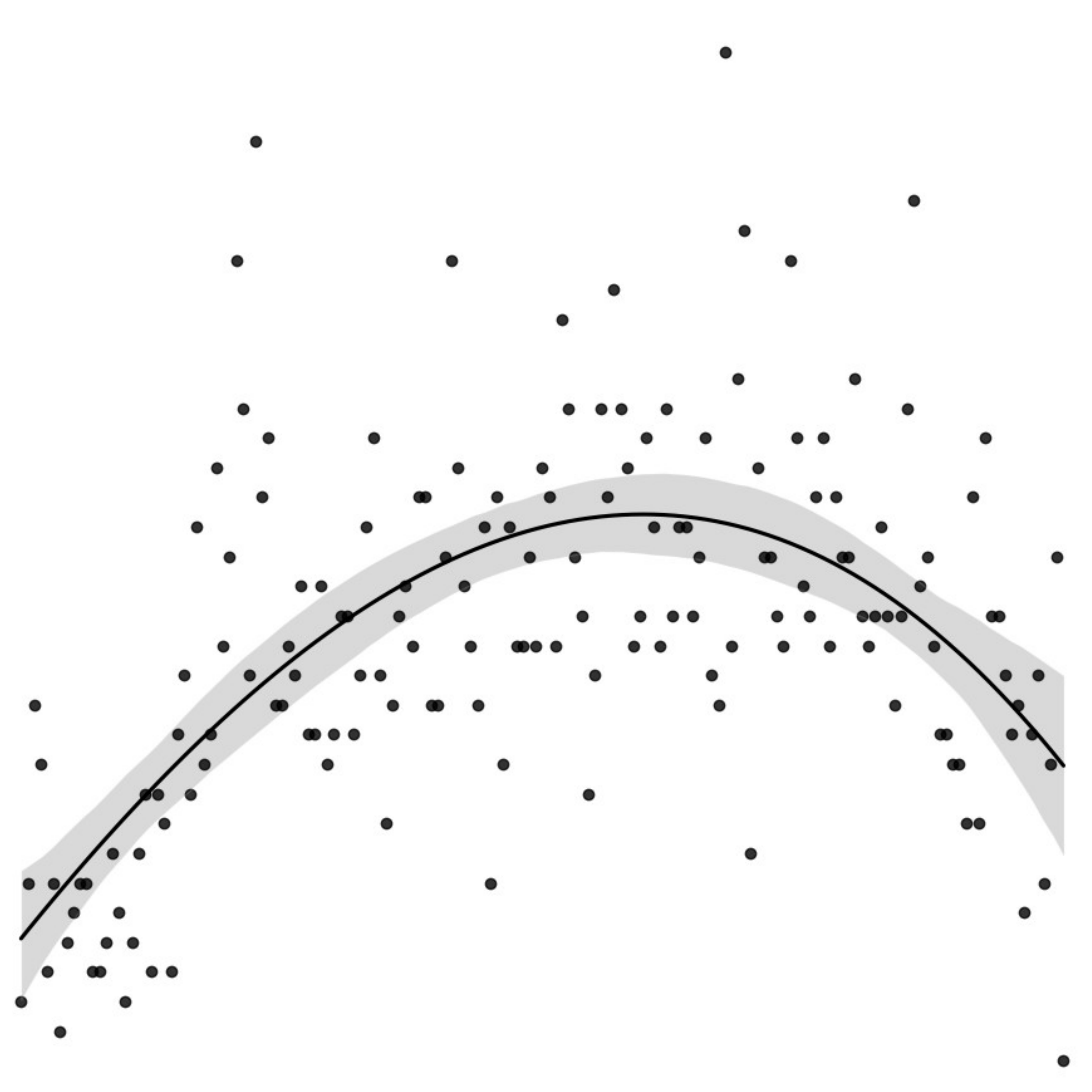}}
	\end{minipage} \\ \hline
T7    & Project managers’ responsibility and contract management & 905 (6.51\%)                                                                &       \begin{minipage}[b]{0.115\columnwidth}
		\centering
		\raisebox{-.5\height}{\includegraphics[width=\linewidth]{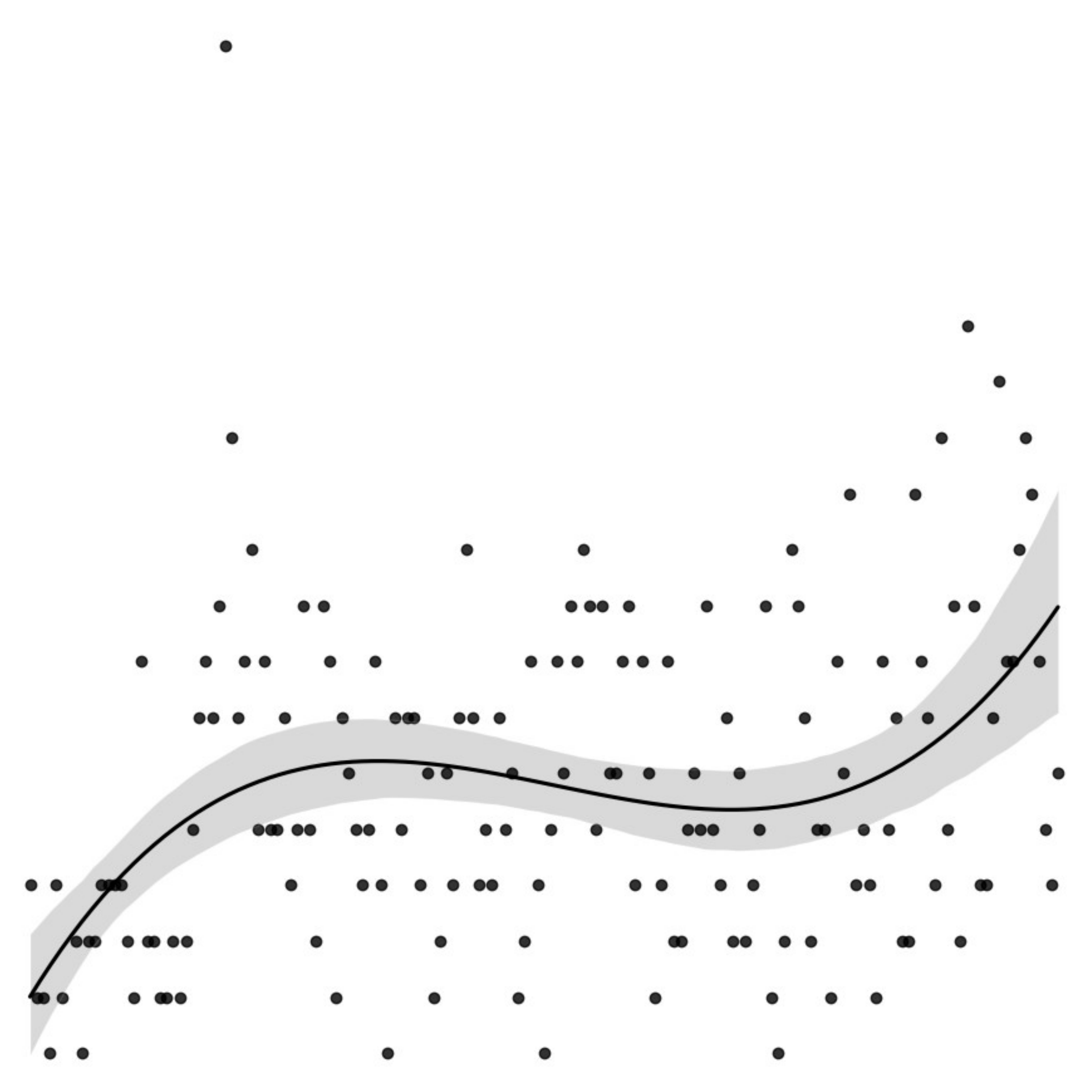}}
	\end{minipage} \\ \hline
T8  & Entities, value Objects, and aggregates in DDD    & 2889 (20.78\%) &  \begin{minipage}[b]{0.115\columnwidth}
		\centering
		\raisebox{-.5\height}{\includegraphics[width=\linewidth]{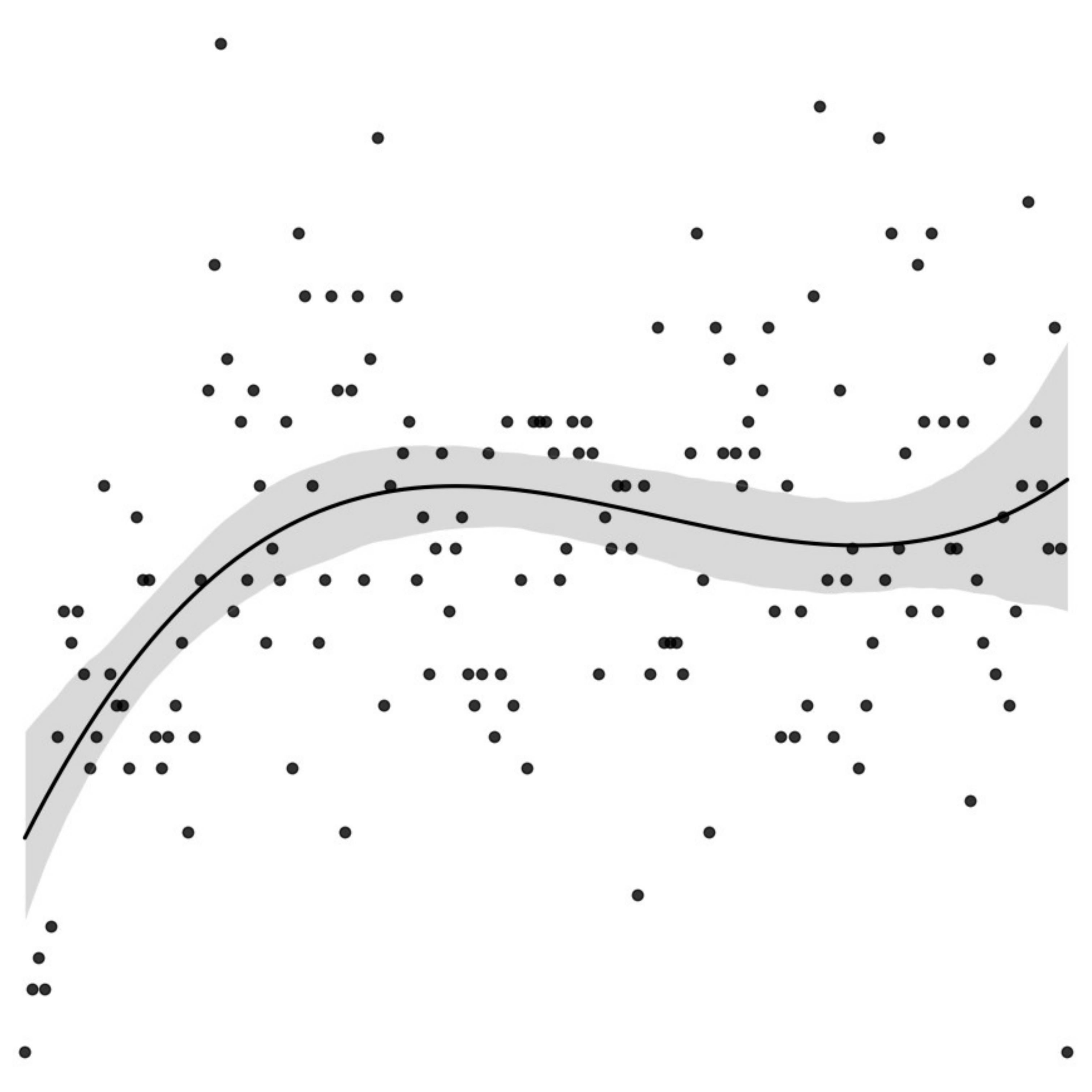}}
	\end{minipage} \\ \hline
T9  & Tools and plugins in agile software development   & 2833 (20.38\%) &  \begin{minipage}[b]{0.115\columnwidth}
		\centering
		\raisebox{-.5\height}{\includegraphics[width=\linewidth]{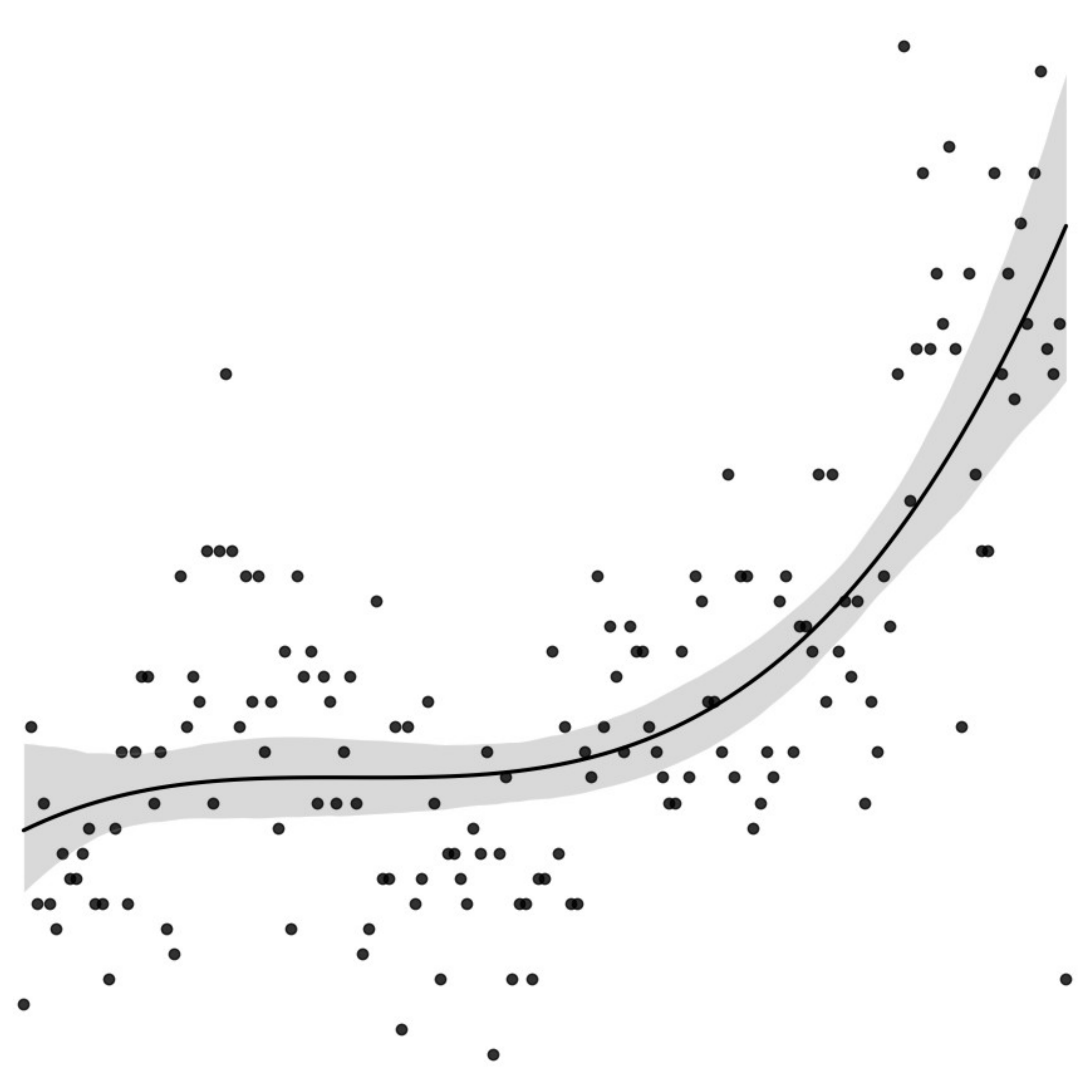}}
	\end{minipage} \\ \hline
T10 & Software development methodology concepts         & 1836 (13.21\%) &  \begin{minipage}[b]{0.115\columnwidth}
		\centering
		\raisebox{-.5\height}{\includegraphics[width=\linewidth]{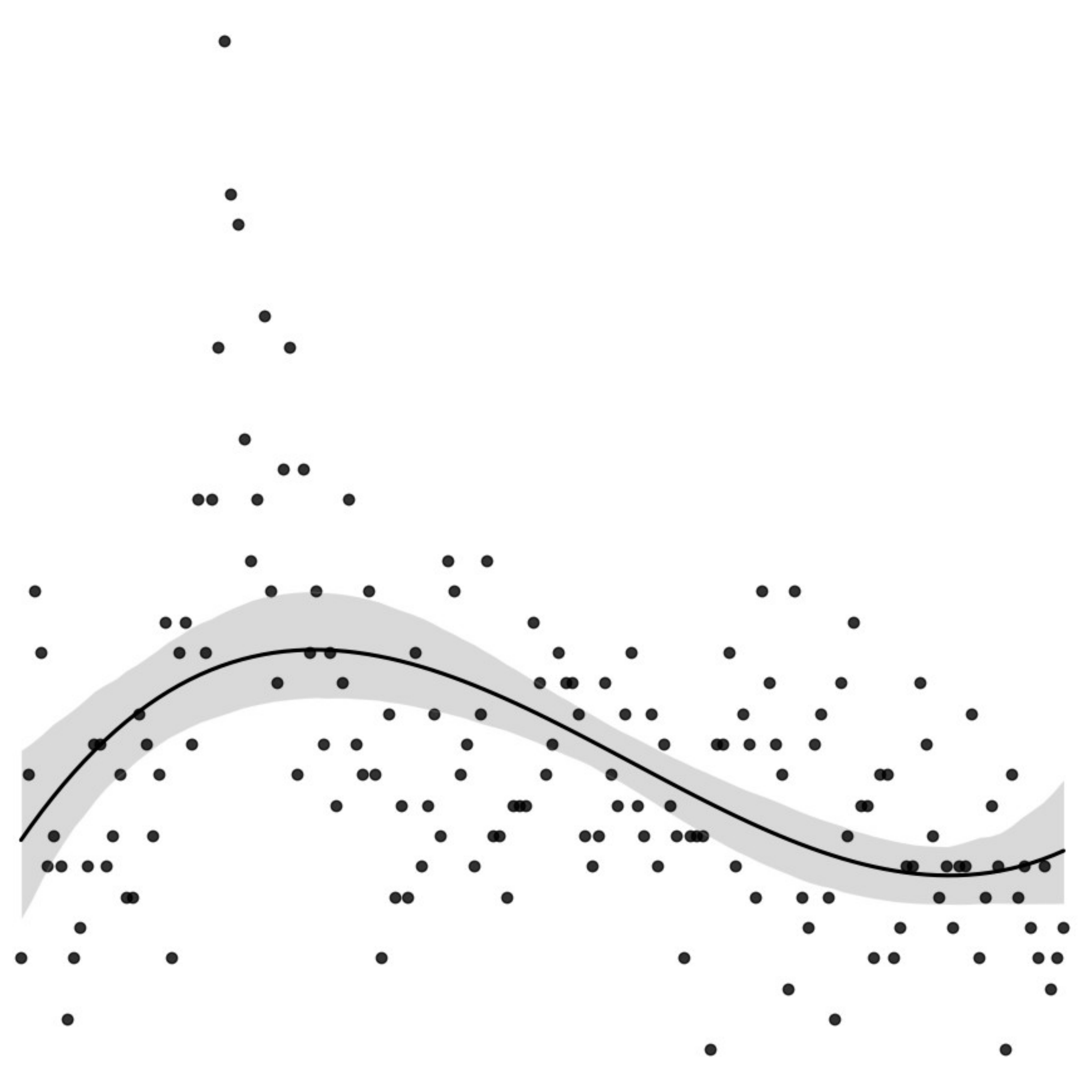}}
	\end{minipage} \\ \hline
T11 & Meetings in agile team                            & 499 (3.59\%)   &  \begin{minipage}[b]{0.115\columnwidth}
		\centering
		\raisebox{-.5\height}{\includegraphics[width=\linewidth]{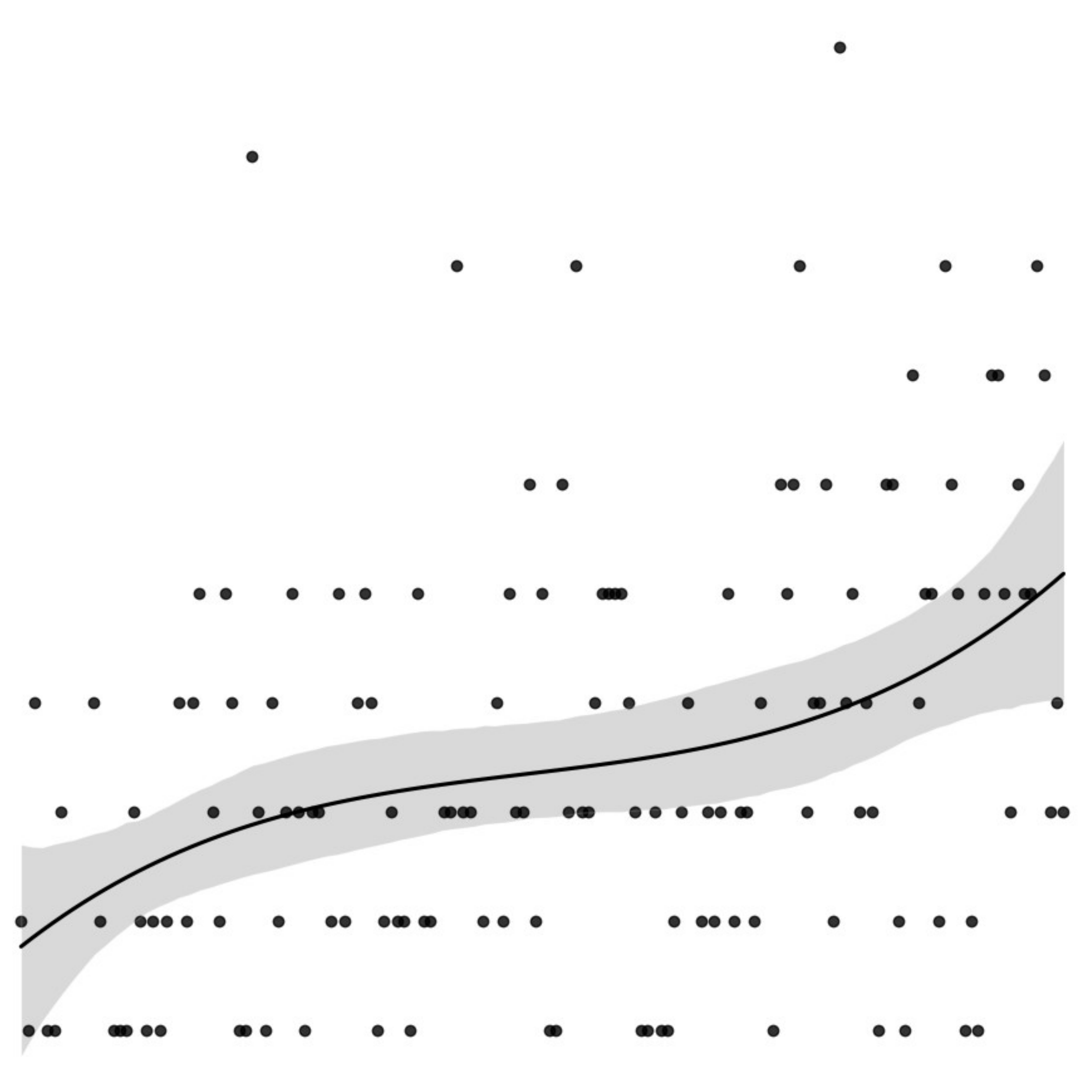}}
	\end{minipage} \\ \hline
T12 & Task management in Kanban Board                   & 2935 (21.11\%) &  \begin{minipage}[b]{0.115\columnwidth}
		\centering
		\raisebox{-.5\height}{\includegraphics[width=\linewidth]{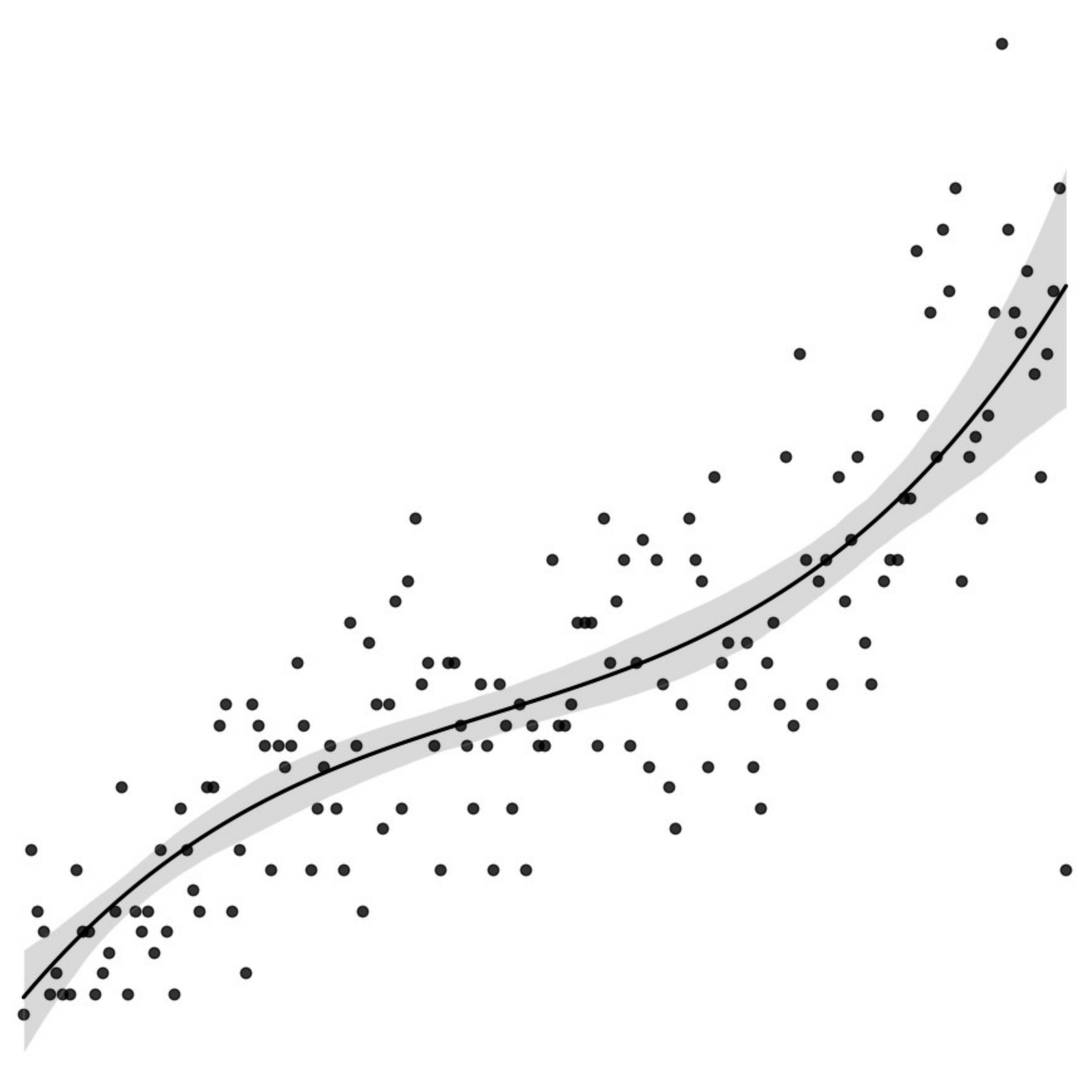}}
	\end{minipage} \\ \hline
T13 & Software testing                                  & 1479 (10.64\%) &  \begin{minipage}[b]{0.115\columnwidth}
		\centering
		\raisebox{-.5\height}{\includegraphics[width=\linewidth]{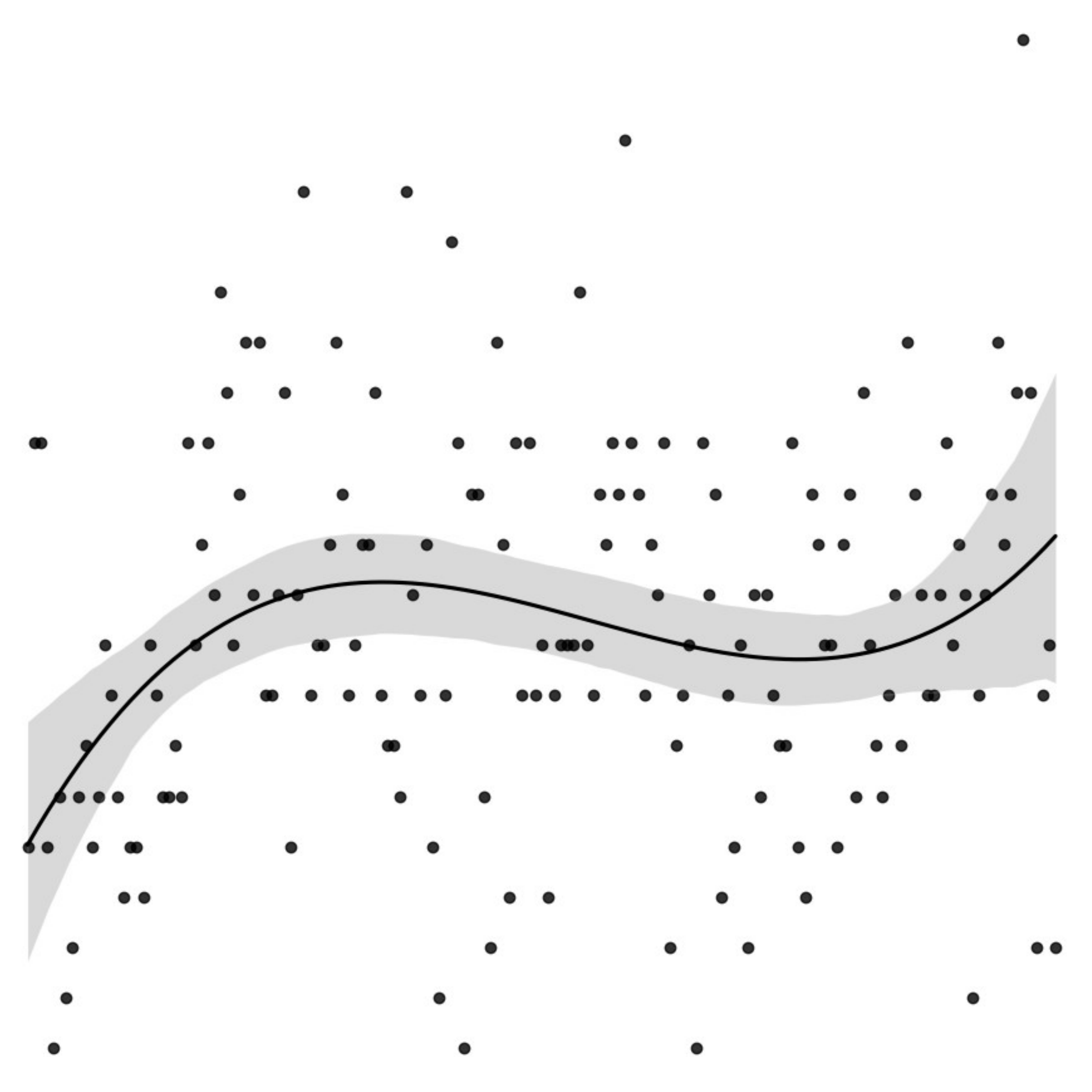}}
	\end{minipage} \\ \hline
T14 & Story estimation in Scrum sprint                  & 2964 (21.32\%) &  \begin{minipage}[b]{0.115\columnwidth}
		\centering
		\raisebox{-.5\height}{\includegraphics[width=\linewidth]{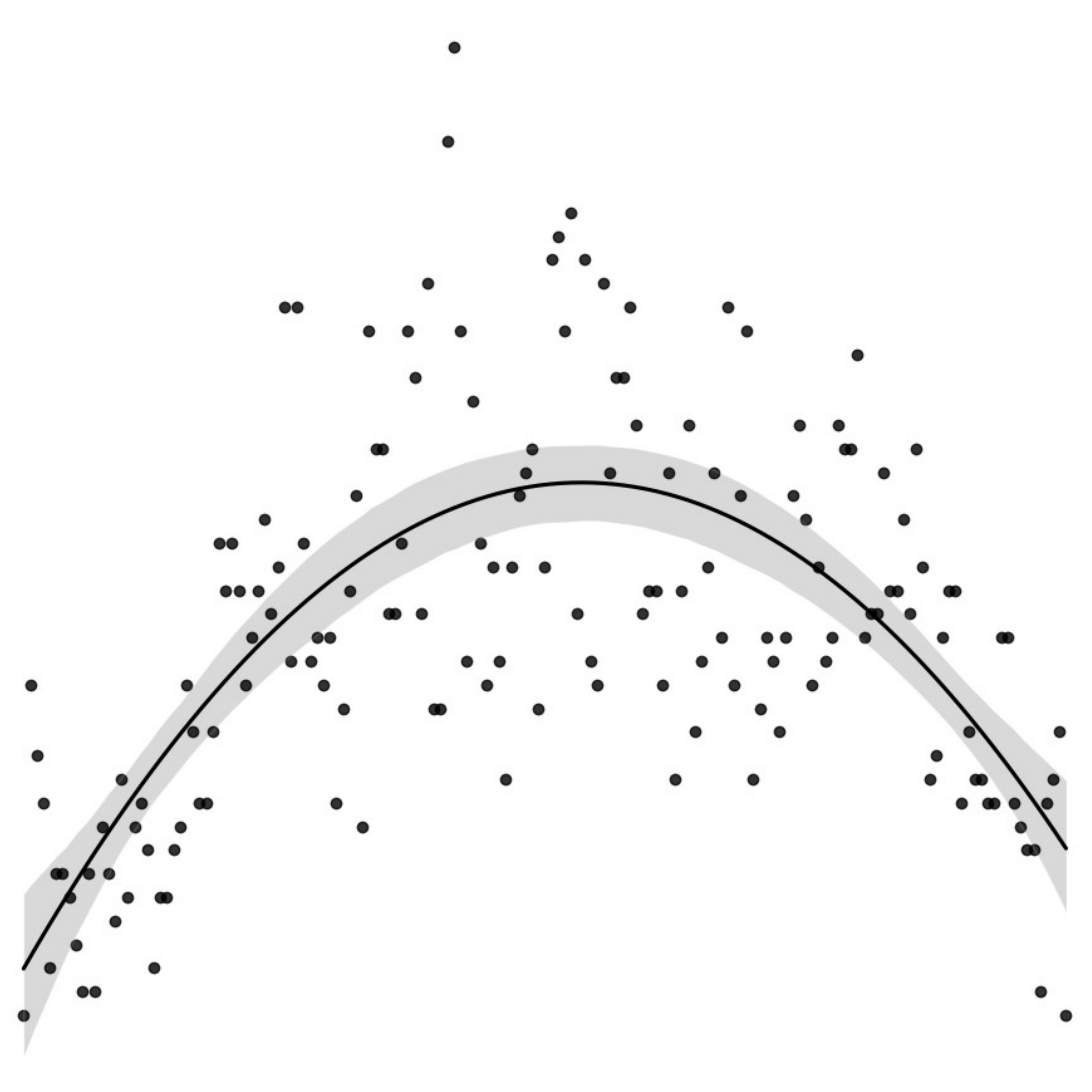}}
	\end{minipage} \\ \hline
T15 & Software design and requirements                  & 3561 (25.61\%) &  \begin{minipage}[b]{0.115\columnwidth}
		\centering
		\raisebox{-.5\height}{\includegraphics[width=\linewidth]{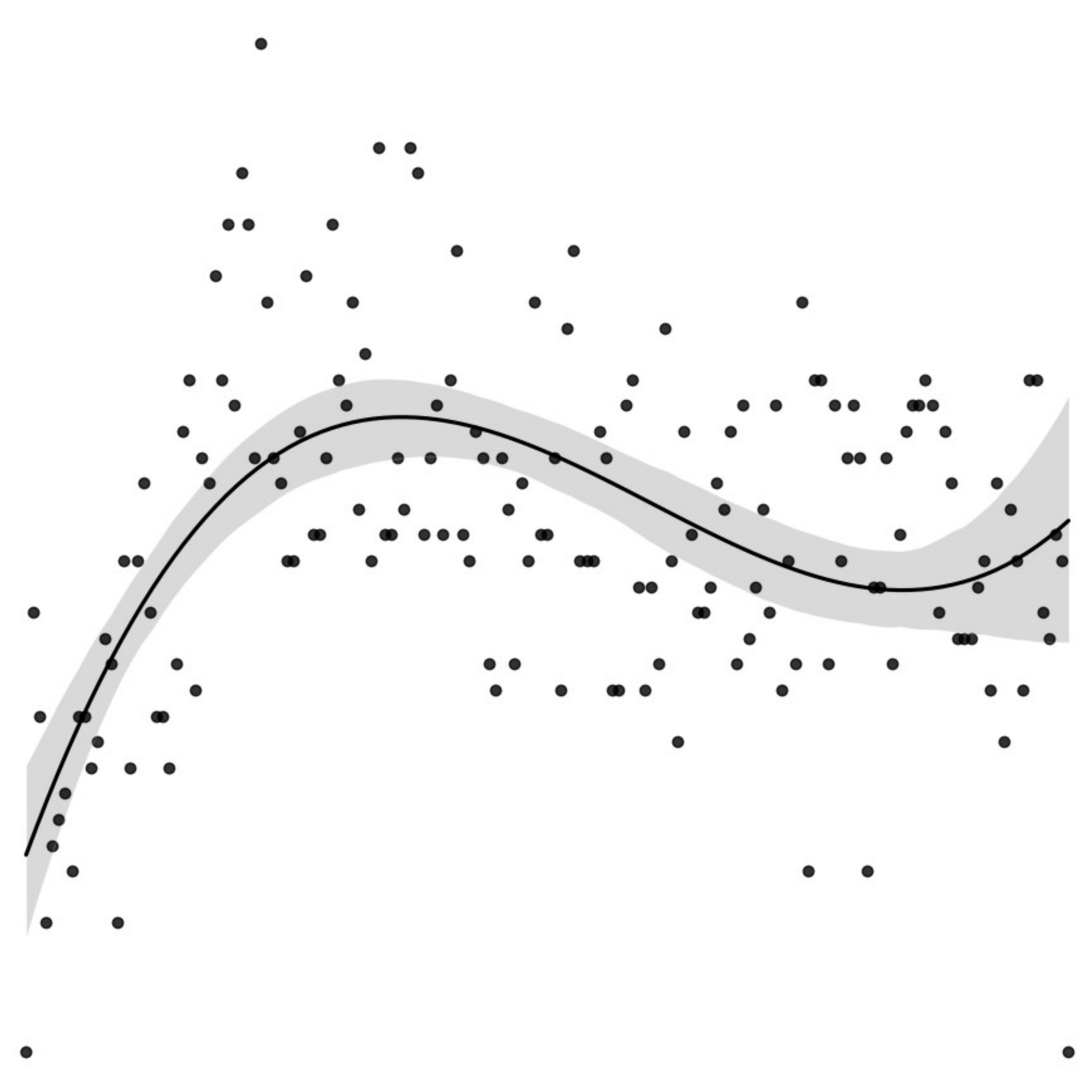}}
	\end{minipage} \\ \hline
\end{tabular}%
}
\end{table}

\textbf{Continues integration, build, and deployment (T1).} 
We labeled developer topics about continuous integration (CI), continuous deployment (CD), and automated pipeline building with tools like Jenkins, Azure DevOps, and GitHub Actions as “T1". CI and CD are essential components of DevOps development and are favored by agile teams. When examining developer posts on SO and SE, we found that many practitioners and software developers struggle to adopt CI, CD, and other automated practices (\textit{ e.g. VSTS Branching Strategy \& CI/CD pipelines- SO post\#50694511)}), managing pull requests in workflows like \textit{(how should we handle pull requests into hotfix/* branches?"- SO post \#39021924)}  and controlling CI pipelines \textit{(e.g. Run pipeline stages only after the previous stage is completed - GitLab CI- SO post \#62919867)}. 

\textbf{Events, bounded contexts and Microservices in DDD (T2).} The Code “T2" is assigned to the developers' topic in the Q\&A websites that discuss the implementation issues of domain events, integration events, bounded contexts, and microservices in Domain-Driven Design (DDD). Domain events are useful for integrating multiple microservices or bounded contexts by providing a messaging-style communication channel. For example, developers asked questions like \textit{(How to fund transfer between bank accounts DDD style with EventStore?- SO post \#54967846)}, \textit{(CQRS + Microservices Handling event rollback- SO post \#46190467)}  and  \textit{(Choreography Sagas in DDD - Chain of Integration Events?- SO post \#63597403)}.

\textbf{Domain model, design patterns, and layers in DDD (T3).} The code “T3" is assigned to developers' topics that focus on the domain model, design patterns, and different layers such as the user interface, application layer, domain layer, and infrastructure layer in DDD. Some example developers' posts of T3 are \textit{(DDD - How to reuse code in application layer?- SO post \#65114332)}, \textit{(Placing entity framework models in the infrastructure layer in onion architecture- SO post \#65474540)}  and \textit{(Could REST API be considered a presentation layer in DDD?- SESE post \#318186)}.

\textbf{DevOps automation tools (T4).} The code “T4" relates to developers discussing technical and configuration matters around DevOps automation tools. Analyzing these discussions, Docker and Kubernetes are often mentioned as applications that aid in automating software development for DevOps teams. Examples include questions like: \textit{(How to run docker containers on different machines- SO post \#34938674)} and \textit{(RancherOS + K8s On a single physical machine with multiple nodes- SO post \#58235277)}.

\textbf{Project, team, and time management (T5).} The code “T5" refers to developers' discussions about project management, teamwork, and time management. Common issues on SO and SE platforms include work in progress (WIP), such as \textit{(Team consistently exceeding their working in progress (WIP) limit- SEPM post \#14049)}, setting development standards \textit{(e.g. Sending out a 'Request for Comment' when establishing a new guideline- SESE post \#56753)} and adopting new technologies in organizations \textit{(e.g. Approach to encouraging organizational adoption of new web dev tools- SESE post \#279820)} . 

\textbf{Team role and responsibilities in Scrum (T6).} The code “T6" pertains to developers' discussions about the roles and responsibilities of Scrum team members, such as the Scrum Master, Product Owner, and development team. Analyzing topics on SO and SE platforms, it is evident that there is confusion regarding the relationship between the Scrum Master and Product Owner, as seen in questions: \textit{ (e.g. In Scrum, is a Scrum Master position higher than a Product Owner?- SEPM post \#15731)}  and \textit{(Can Product Owner + Scrum Master (Team/Kanban Lead) provide combined leadership for a team?- SEPM post \#17287}). Other posts explore the Product Owner's role, for example, \textit{(Product Owner role in Scrum- SEPM post \#11613)}, or address development teams in different locations, such as \textit{(A Scrum Master is working with a development team that has members in different physical locations- SEPM post \#26223)}.

\textbf{Project managers’ responsibility and contract management (T7).} The code “T7" refers to developers' discussions about the responsibilities of management roles and the significance of contracts in project assurance. Analyzing topics on SO and SE platforms, it is clear that many practitioners talk about project managers' duties in posts such as \textit{(Agile management roles- SEPM post \#6769)} and \textit{(Financial indicators of a project and a project manager's efficiency- SEPM post \#30733)}. Additionally, developers often discuss contract management in posts  \textit{(e.g. How can a PM manage the contract aspects for Scrum projects with outcome-based pricing? SEPM post \#29636)} and \textit{(How is a software development contract concluded? SEPM post \#30284)}.

\textbf{Entities, Value Objects, and aggregates in DDD (T8).} In DDD, entities have a unique identity and combine data and behavior, like users or products. Value objects describe characteristics without a unique identity. Aggregates are collections of related entities and value objects, simplifying the management of complex DDD systems. Each aggregate has an entity, known as the aggregate root, which controls access rights for objects outside the aggregate. The code “T8" is given to developers' posts on SO and SE platforms that discuss aggregates and related concepts. During the analysis, it is noted that developers frequently ask about concepts \textit{(e.g. DDD, Aggregate roots and Entities- SO post \#46661591)}  and practices such as \textit{(DDD: is it correct for a root aggregate to hold a reference to another root aggregate?- SESE post \#328571)}.

\textbf{Tools and plugins in agile software development (T9).} The code “T9" is given to posts focusing on commonly used tools and plugins in agile software development approaches. For instance, developers ask about suitable IDEs \textit{(e.g. Javascript IDE for agile development- SO post \#7090013)}, PHP open-source tools such as \textit{(PHP Open Source Tools for Agile Development- SESE post \#86883)}, or cross-platform project management tools like \textit{(Any agile free cross-platform project management/ALM tools with Mylyn integration out there?- SO post  \#3863050)}.

\textbf{Software development methodology concepts (T10).} The code “T10" is given to developers' discussions on SO and SE platforms that focus on concepts of software development methodologies, including Extreme Programming, Waterfall, and Scrum. For example, developers ask questions like \textit{(What's the relationship between SDLC and methodologies like XP, RAD, Scrum, etc.?- SEPM post \#15378)} and  \textit{(Software development methodology difference- SO post \#6147478)}.

\textbf{Meetings in agile team (T11).} The code “T11" is given to posts focusing on common meetings agile teams hold during product development. These meetings include but are not limited to, daily stand-ups \textit{(e.g. Should we cancel the daily stand-Up if we have another meeting during the day?- SEPM post \#20796 )} and \textit{(Can a daily scrum meeting be replaced by a status email?- SESE post \#210111)}, sprint reviews, and sprint retrospectives \textit{(e.g. What's the ideal length of the sprint review/retrospective based on the length of the iteration?- SESE post  \#71926 )}.

\textbf{Task management in Kanban board (T12).} Kanban board offers a clearer view of project workflow by visually displaying the work of all team members in software development. As a result, we assigned the code “T12" to practitioners' discussion topics that address problems or issues related to Kanban Board. Developers need to efficiently organize and manage tasks on the Kanban board, as seen in questions like \textit{(Why in JIRA my field Resolution is labeled as Unresolved when the status is Resolved?- SEPM post 
 \#22987)}  and \textit{(Why are items for other users on the TFS current iteration task board grey?- SO post \#18450086)}.

\textbf{Software testing (T13).} Software testing helps detect bugs or errors early and offers solutions before delivery, improving software application quality and increasing user satisfaction. Incorporating software testing in development approaches ensures security, reliability, and high performance. We assigned the code “T13" to developers' posts focusing on software testing-related issues, including unit testing, integration testing, acceptance testing, and test-driven development (TDD). Examples of developer discussions include: \textit{(Unit testing - Practicality: Should it pass or fail all or most of the time?- SO post \#33205413)}  and \textit{(test-driven development - Who should write the tests?- SESE post \#35610)}.

\textbf{Story estimation in Scrum sprint (T14).} Story estimation helps the Product Owner gauge the effort required for each user story to complete the software project on time and within budget. Each team member provides their perspective on the work during story estimation. We assigned the code “T14" to practitioners' discussion posts on selected Q\&A websites that address issues or problems related to story estimation. Common story estimation questions include when to estimate, such as \textit{(Should a scrum team estimate time for the user stories during Sprint Planning, or before it?- SESE post \#270689)}  and \textit{(When is it time to do poker planning - during Story Time or during Sprint Planning?- SEPM post \#15545)}, and how to use story points for estimation, e.g.  \textit{(Measuring scale for story points in Scrum framework- SO post \#9017428)}.

\textbf{Software design and requirements (T15).} Detailed, clear, and accurate software requirements help designers, developers, and testers better understand product functionalities and develop software applications that lead to higher user satisfaction. Improved software requirements elicitation, specification, and validation can also streamline the design phase, resulting in higher-quality software applications. We assigned the code “T15" to developers' discussion posts on SO and SE platforms where practitioners discuss challenges related to requirements, design, or both. Numerous posts cover software design and requirements, such as \textit{(Requirements with units in Software Design Document- SESE post \#332967)}. Unified Modeling Language (UML) is also frequently mentioned in the design and requirement analysis phases, e.g. \textit{(Include Use Case Diagram (UML)- SO post \#31759144)}. 

In summary, our analysis of software development approaches related to topics on selected Q\&A repositories provides valuable insights into the most frequently discussed areas. The top three highly asked topics include “T5: Project, team and time management," “T3: Domain model, design patterns and layers in DDD," and “T15: Software design and requirement." These results demonstrate the importance of effective project and team management, a strong understanding of domain-driven design, and the crucial role of software design and requirements in addressing developers' concerns and challenges within the software development process.

\begin{tcolorbox}[colback=gray!5!white,colframe=gray!75!black,title=Key Findings of RQ2]
\textbf{Finding 2}: The top three highly discussed topics are “T5: Project, team, and time management," “T3: Domain model, design patterns, and layers in DDD," and “T15: Software design and requirement", emphasizing the importance of these aspects in addressing developers' concerns and challenges. 

\textbf{Finding 3}: Practitioners often discuss CI/CD, DevOps tools, and Agile methodologies like Scrum, Kanban, and Extreme Programming. This suggests that developers actively engage with these methods and need guidance to effectively incorporate them into their workflows.

\textbf{Finding 4}: Other notable topics include software testing, story estimation in Scrum, and Kanban task management- highlighting the demand for effective strategies to ensure quality, precise estimates, and improved task handling in software projects.
\end{tcolorbox}

\subsection{RQ2.1: Popular and Difficult Topics} \label{Sec: Popular and Difficult Topics}
\begin{figure*}[h]
  \centering
  \includegraphics[width=\linewidth]{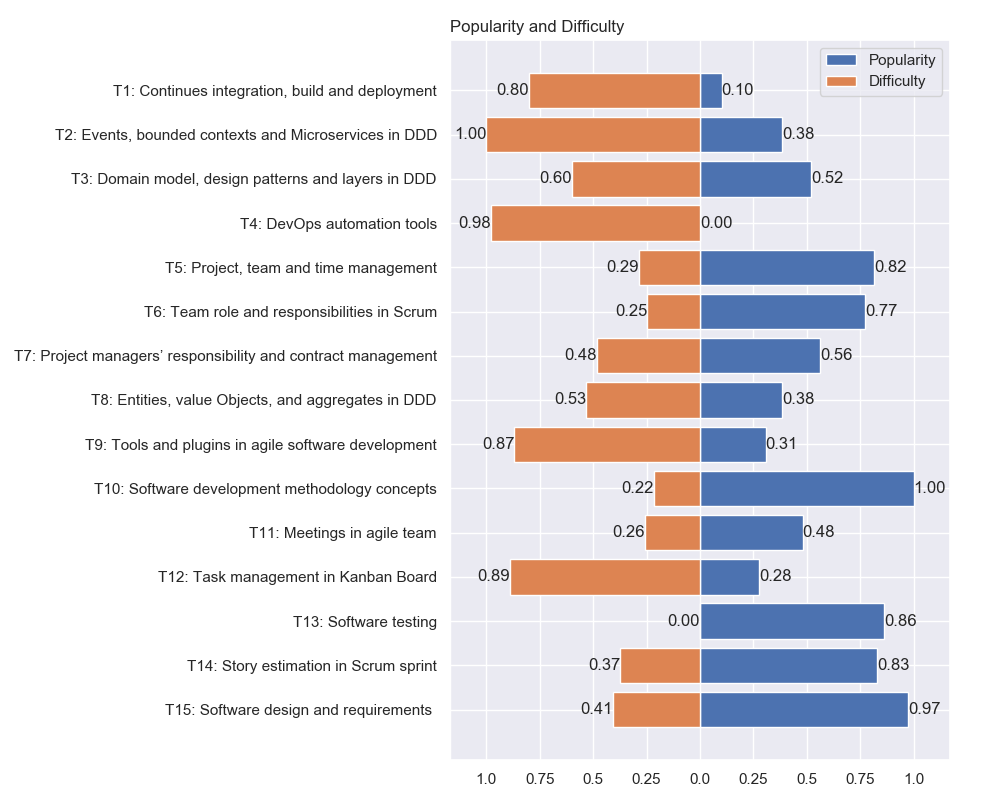}
  \caption{Popularity and difficulty of topics}
  \label{Fig.popularity_difficulty}
\end{figure*}

We further analyzed the topics discussed in Section \ref{RQ2: Software Development Approaches Topics} (RQ2) to identify the most popular and difficult topics to address. 
As shown in Figure \ref{Fig.popularity_difficulty}, we applied the min-max normalization method (Equation 5) to adjust the values of popularity and difficulty. The full statistical metrics for popularity and difficulty can be found in \cite{peng2022topics} and the details are provided in Section \ref{sec:research-method-questions} (RQ2.1). As follows, we explore the popular and difficult topics that can help practitioners focus their efforts on addressing the most pressing issues and improve their understanding of complex concepts.

\textbf{Topic popularity.} The top three most popular topics on the SO and SE platforms have been identified using min-max normalization. These topics are Software development methodology concepts (T10), Software design and requirements (T15), and Software testing (T13) (See Figure \ref{Fig.popularity_difficulty}). Although T10 is the most popular topic, it ranks 11th in the frequency distribution of all topics, as depicted in Figure \ref{Fig.distribution}. We used Kendall's Tau correlation test \citep{kendall1970rank} to evaluate the relationship between topic popularity and frequency distribution. The resulting Tau value is 0.028, with a probability of tau = 0 (no association) and a p-value of 0.923. This suggests no significant correlation exists between the topic's popularity (Figure \ref{Fig.popularity_difficulty}) and its frequency distribution (Figure \ref{Fig.distribution}).

It demonstrates that the frequency of a topic does not accurately represent its popularity- the number of questions on a topic cannot fully capture user activity on a Q\&A website. By understanding these findings, software practitioners can gain insight into the prevailing topics in the field of software development approaches. This information can prove valuable when selecting the most suitable software development approach.

\begin{figure*}[h]
  \centering
  \includegraphics[width=0.75\linewidth]{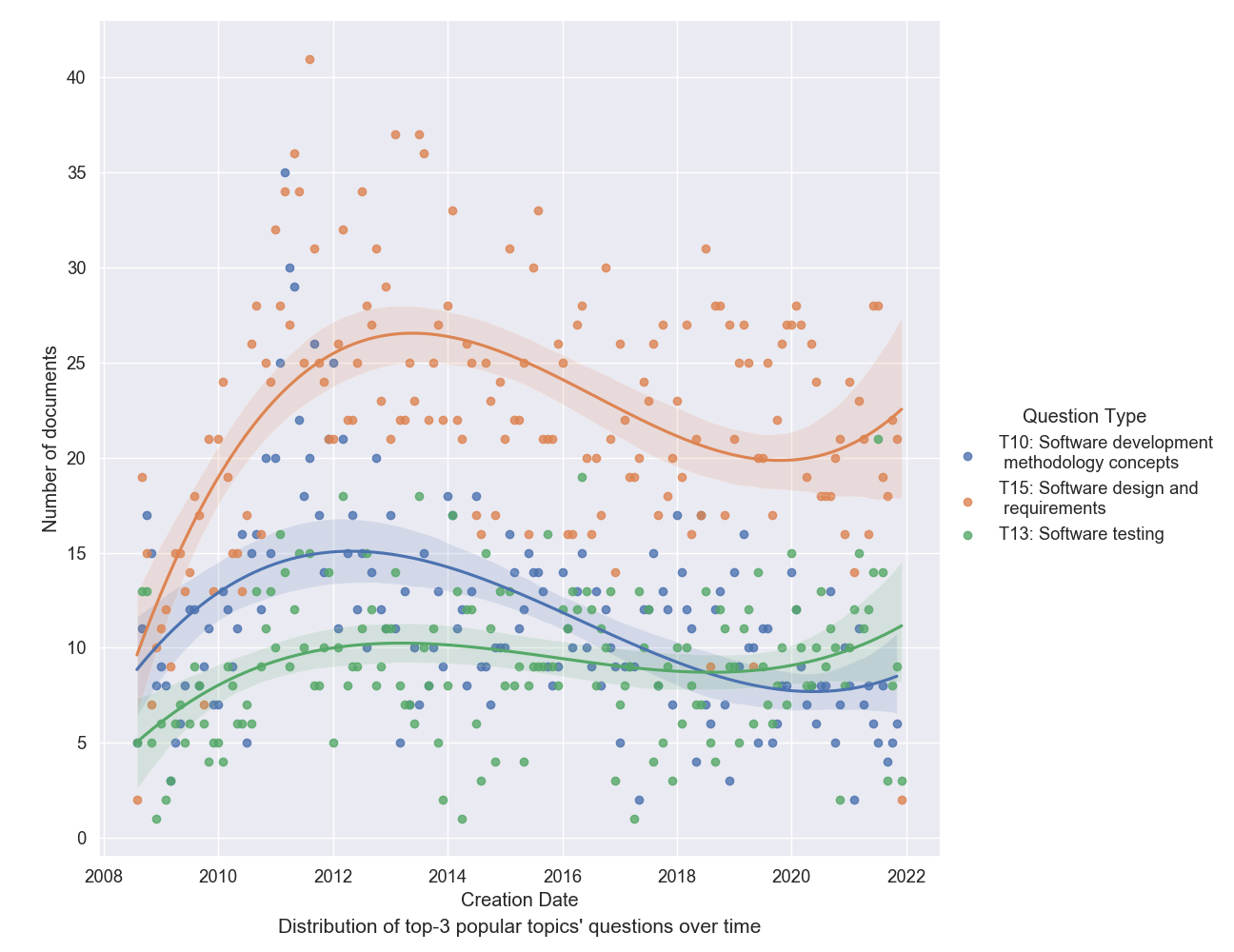}
  \caption{Distribution of top-3 popular topics’ questions over time}
  \label{Fig.top-3-popularity}
\end{figure*}

\begin{figure*}[h]
  \centering
  \includegraphics[width=0.75\linewidth]{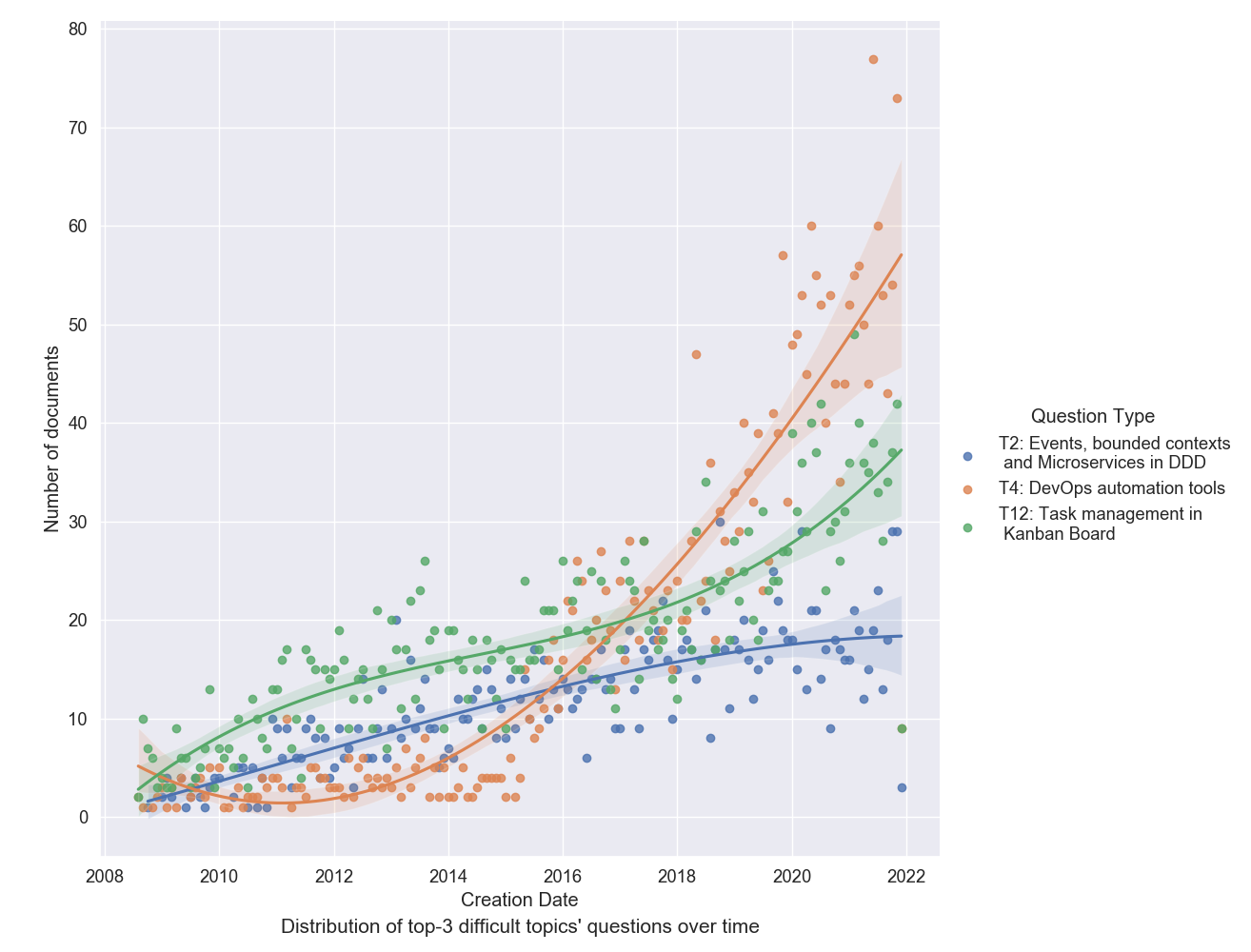}
  \caption{Distribution of top-3 difficult topics’ questions over time}
  \label{Fig.top-3-difficulty}
\end{figure*}

\textbf{Topic difficulty.} In the dataset, the top three most difficult topics for developers were identified as Events, bounded contexts, and Microservices in DDD (T2), DevOps automation tools (T4), and Task management in Kanban Board (T12) (see Figure \ref{Fig.popularity_difficulty}). From Figure \ref{Fig.top-3-difficulty}, it is evident that the number of posts on any of the top three most difficult topics exhibited an upward trend. For instance, for DevOps automation tools (T4), the developers' questions increased significantly (See Figure \ref{Fig.top-3-difficulty}). This information is valuable for software practitioners to continuously monitor developers' posts on Q\&A websites and provide timely solutions or remedies for questions regarding emerging trends. In contrast, the number of questions for the top three most popular developers' topics did not increase, with a downward trend visible in Figure \ref{Fig.top-3-popularity}. Using Kendall's Tau correlation test \citep{kendall1970rank}, we discovered a strong correlation (tau = -0.600, p-value = 0.001) between topic popularity and difficulty (popularity decreased as difficulty increased).

These findings indicate that many questions related to these topics remain unaddressed, and software development experts should devote more attention and effort to providing solutions for questions associated with these subjects (topics). To improve software application quality, developers should address unanswered questions in difficult areas and should actively monitor and incorporate these topics to advance existing development approaches.

\begin{tcolorbox}[colback=gray!5!white,colframe=gray!75!black,title=Key Findings of RQ2.1]
\textbf{Finding 5}: The top three most popular topics are Software development methodology concepts (T10), Software design and requirements (T15), and Software testing (T13).  No strong correlation exists between frequency distribution (Figure \ref{Fig.distribution}) and popularity (Figure \ref{Fig.popularity_difficulty}), indicating that question count doesn't fully represent user activity on Q\&A websites.

\textbf{Finding 6}: The top three most difficult topics for developers are Events, bounded contexts, and Microservices in DDD (T2), DevOps automation tools (T4), and Task management in Kanban Board (T12). A strong negative correlation between popularity and difficulty indicates that difficult topics are less popular.

\textbf{Finding 7}:
Many questions on difficult topics go unanswered, indicating practitioners should focus on addressing them. Providing solutions for these areas can enhance software quality and progress development approaches.
\end{tcolorbox}

\subsection{RQ3: Software development approaches challenges}

\begin{figure*}[h]
  \centering
  \includegraphics[width=1\linewidth]{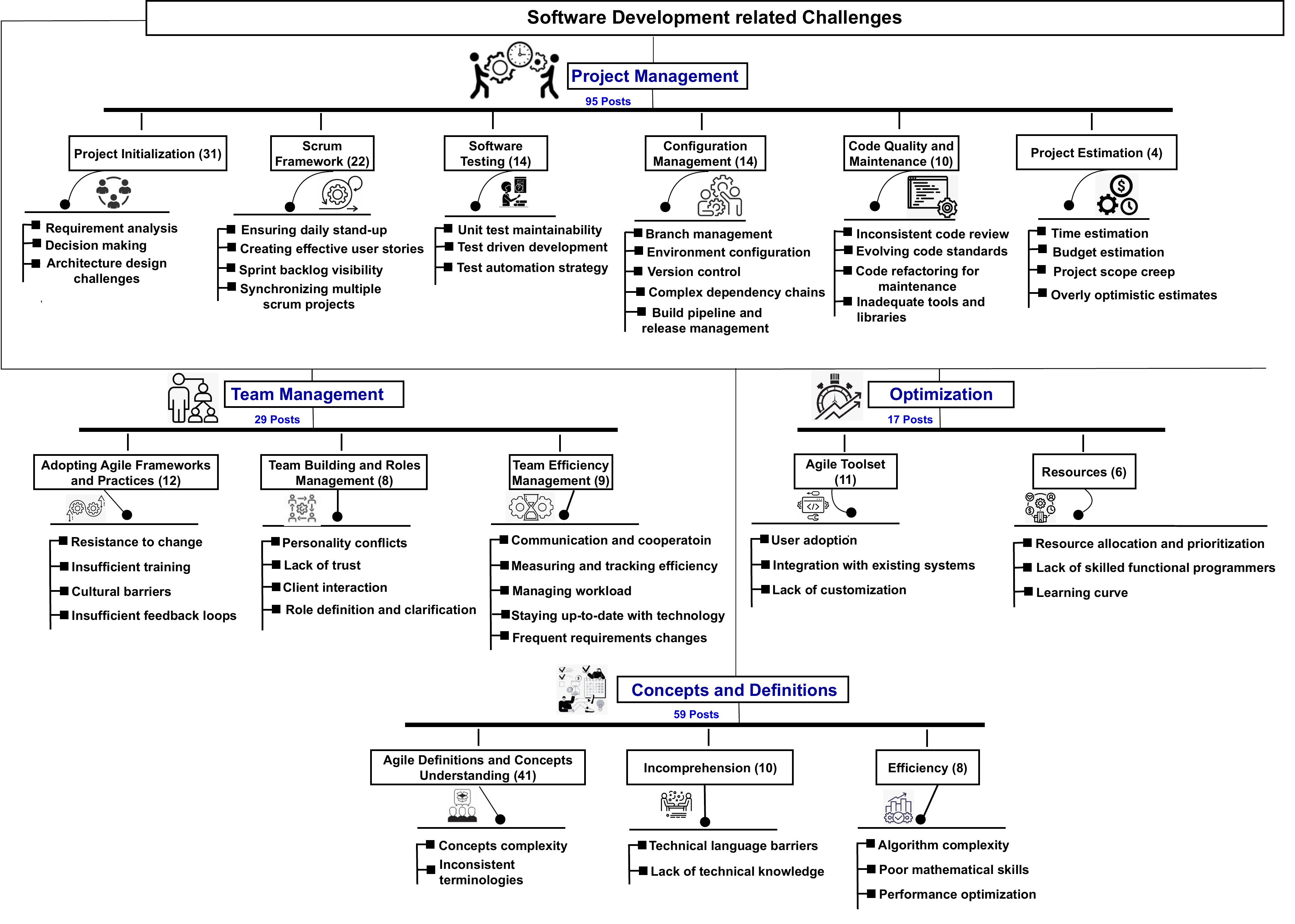}
  \caption{Mapping of identified challenges}
  \label{Fig.challenges-mapping}
\end{figure*}

This paper also aims to identify critical challenges practitioners encounter when dealing with software development approaches. To achieve this, we conducted a detailed qualitative analysis, as described in Section \ref{sec:research-method-questions} (RQ3), on the top 200 highly ranked developers' posts on the SO and SE platforms, sorted by AMS \citep{bajaj2014mining}. We identified 49 challenges, which were mapped across 14 sub-themes and 4 high-level themes (categories), as illustrated in Figure \ref{Fig.challenges-mapping}. 
The summary of the high-level themes (categories) is as follows:
\subsubsection{Project management challenges}
We analyzed developers' discussions on selected Q\&A platforms, focusing on the project management challenges they encounter when adopting software development approaches and practices. This category comprises the most significant portion of the dataset, including 95 out of 200 top-ranked developers' posts. These posts highlight practitioners' questions for effective and efficient solutions in software project management. Examining the data from these 95 posts, we identified 22 project management relevant challenges classified across six sub-themes (See Figure \ref{Fig.challenges-mapping}).


In project management, practitioners face challenges related to project initiation activities such as requirement analysis, decision-making, and architecture design. For example,  \textit{(How can I get things right at the beginning of a software project?- SESE post \#324082)}. These initial steps provide a fundamental structure; if they go wrong, they can lead to project failure. Software practitioners should provide guidelines with examples to help in initiating and executing project activities more efficiently.

A common sub-theme of developers' questions in this category relates to challenges when adopting the Scrum framework, such as daily stand-up, user stories, sprint backlog, and project synchronization. For example, \textit{(Developers wonder if bug-fixing tasks should be assigned story points in Scrum- SESE post \#162145)}. Similarly, developers often seek technical solutions for software testing (sub-theme), such as unit testing and Test-Driven Development \textit{(Is unit testing or test-driven development worthwhile?- SESE post \#140156)}, on Q\&A platforms.

We also identified configuration management (sub-theme) challenges, e.g., branch management, environment configuration, version control, complex dependency change, build pipelines, and release management (See Figure \ref{Fig.challenges-mapping}). For instance, \textit{(How do you handle integrating code from multiple branches/developers each sprint?- SESE post \#372716)}. Furthermore, developers' discussions revealed challenges relevant to code quality and maintenance such as \textit{(I’ve
inherited 200K lines of spaghetti code – what now? SESE post \#155488)} and project estimation \textit{(What can I do to get better at estimating how long projects are going to take?- SESE post \#39411)}.

In summary, practitioners face a variety of project management challenges when adopting different approaches and practices for software development. The main areas of concern include project initialization, Scrum framework adoption, software testing, configuration management, code quality, and project estimation. These insights are vital for software practitioners to improve their methodologies and better support development activities. Furthermore, researchers can use these findings to develop novel software development approaches that address current trends and challenges faced by practitioners in the field.

\subsubsection{Team management challenges}
This category includes developers' posts from the selected Q\&A platforms that discuss challenges to effectively managing a team in software development approaches. We identified a total of  13 primary challenges, which were subsequently grouped into 3 subcategories (See Figure \ref{Fig.challenges-mapping}).

Upon careful examination of developers' posts, we noticed that development teams encounter various challenges when adopting agile frameworks and implementing their practices. For instance, \textit{(How can we make Agile enjoyable for developers that like to personally, independently own large chunks from start to finish- SESE post \#80751)}, specifically when shifting the team from Waterfall to Agile.

Developers also encountered challenges relevant to team building and roles management, such as \textit{(Why can't the Scrum Master and the project manager be the same person?- SEPM post \#4707)}. Another example of developers' posts in this category is seeking approaches to improve team efficiency, e.g. \textit{(How can we reduce downtime at the end of an iteration? SESE post \#66708)}.

We further noticed five core challenges related to team efficiency management (See Figure \ref{Fig.challenges-mapping}), for example, \textit{(How to measure software development performance?- SO post \#1168131)}, \textit{(How to tell whether your programmers are under-performing?- SESE post \#177167)} and \textit{(Product owner and/or scrum master in performance review of developers- SESE post \#439092)}.

 It can be concluded that developers face several challenges when managing teams in software development approaches, particularly when transitioning from one methodology to another, such as from Waterfall to Agile. Key concerns include making Agile enjoyable for developers who prefer to independently own large project portions, managing and defining team roles, and improving team efficiency. Addressing these challenges is crucial for organizations to ensure smooth transitions between development methodologies and to enhance overall team productivity and effectiveness.

\subsubsection{Optimization}
In this category, developers' discussion posts related to challenges focusing on agile toolset and resources. Various posts focuses on Agile tools choice, comparison, and usage, with examples including Azure Web App \textit{(Azure Web App (ASP.NET MVC) becomes cold every ten minutes and takes +10-20s to load- SO post \#50115939)}, Azure DevOps Server \textit{(backup of Azure DevOps repositories - SO post \#62174938)}, Kanban Board \textit{(Where to Find a Desktop Kanban board application?- SEPM post \#829)}, Jenkins \textit{(Is it a good idea to make Ansible and Rundeck work together, or using either one is enough?- SO post \#31152102)}, Visual Studio \textit{(What is Security Development Lifecycle Checks option in Visual Studio?- SO post \#18304632)}, and Virtual Machine \textit{(Thoughts on Development using Virtual Machines- SESE post \#103501)}. Additionally, developers also discussed resources-related problems, for example, lack of skilled functional programmers \textit{(How much functional programming expertise can programmers be expected to have? SESE post- \#209071 )} and \textit{(Is there a software-engineering methodology for functional programming? SO post- \#4852251)}, inadequate tools and libraries \textit{(Where do I find some good examples for DDD?- SO post \#540130)} that can lead to suboptimal project outcomes and hinder overall progress in software development (See Figure \ref{Fig.challenges-mapping}).

These findings are encouraging for both software developers and the research community. Software developers can provide best practices and post-mortem reports on software methodologies to help in understanding the pros and cons of development approaches. Similarly, software researchers can seize this opportunity to develop research studies comparing various software approaches (traditional and Agile) in developing software applications, particularly considering market-based software products.

\subsubsection{Concepts and definitions}
In this category, developers' discussion posts focus on asking general questions in Q\&A platforms to gather community knowledge on the concepts and definitions of software development approaches and practices. We identified 7 relevant challenges, which were further categorized across three sub-themes (See Figure \ref{Fig.challenges-mapping}).

One common sub-theme in the selected posts involves the agile definition and concepts understanding. For example, practitioners sought information on Domain-Driven Design \textit{(Value objects in DDD - Why immutable?- SO post \#4581579)}, Scrum \textit{(What is the difference between Sprint and Iteration in Scrum and length of each Sprint?- SO post \#1227318)}, Test-Driven Development (TDD) \textit{(Why is agile all about the test-driven development (TDD) and not development-driven test?- SESE post \#326485)}, Continuous Integration (CI) \textit{(What is the purpose of a dedicated "Build Server"?- SO post \#1099133)}, Scaled Agile Framework (SAFe), Nexus and Large Scale Scrum (LeSS) \textit{(Scaled scrum/agile frameworks (SAFe vs. Nexus vs. LeSS) comparison - SEPM post \#17441)}.

Incomprehension is also a sub-theme in this category; it revolves around issues in technical language barriers and lack of technical knowledge, such as \textit{(How to sell Agile development to (waterfall) clients?- SESE post \#215562)}  and \textit{(Getting non-programmers to understand the development process- SESE post \#4)} .

Efficiency is the third sub-themes of the concept and definition category, where practitioners seek clarification on how algorithms and mathematics efficiently apply to software development approaches, e.g.,  \textit{(What does mathematics have to do with programming?- SESE post \#136987 )} and \textit{(Is big-O really that relevant when working in industry?- SESE post \#20832)}. Lastly, several posts in this category aim to optimize the performance of software development approaches, e.g., \textit{(Why can't the IT industry deliver large, faultless projects quickly as in other industries?-SESE post \#158640)}.

It indicates that developers actively seek knowledge on various software development approaches, including their concepts, definitions, characteristics, and advantages. They are curious about the relevance of mathematics and algorithms in these approaches. Additionally, developers are interested in optimizing the performance of software development approaches. This highlights the importance of community-driven knowledge sharing and continuous learning in the software development field to address these concerns and enhance overall understanding.
\begin{tcolorbox}[colback=gray!5!white,colframe=gray!75!black,title=Key Findings of RQ2.1]
\textbf{Finding 8}: Practitioners face challenges in various software development approaches, which are grouped into four high-level themes: Project Management, Team Management, Optimization, and Concepts and Definitions.

\textbf{Finding 9}: Most challenges in software development are related to project management (See Figure \ref{Fig.challenges-mapping}). Ineffective project management can lead to delays, budget overruns, and poor-quality software. Thus, it's vital for practitioners to identify and overcome project management challenges for successful software development project completion.

\end{tcolorbox}

%% file: implication.tex
\section{Implications}
\label{sec:implication}
We now summarised the research and industrial implications of the study findings as follows:

\subsection{Research implications}

Examining practitioners' discussions on popular Q\&A sites holds significant value for researchers in enhancing software development approaches and application quality \cite{rosen2016mobile}. The trends observed in this study suggest that researchers should devote more attention to the software development approaches domain by providing comprehensive documentation, case studies, and software post-mortem reports. The strong correlation between topic popularity and difficulty highlights the need to focus on and invest effort into various software development process activities that are relatively difficult to understand and execute. Moreover, the thematic mapping of challenges in software development approaches offers a valuable framework for researchers to explore key issues faced by practitioners. By focusing on these challenges, researchers can design targeted studies to understand the underlying causes and develop effective solutions (tools, frameworks, or techniques) tailored to the specific needs of practitioners.

\subsection{Industrial implications}
The research findings on developers' discussion topics on Q\&A sites can be employed by software practitioners to understand the current state and trends in software development approaches and the difficulties or challenges developers encounter when adopting software methodologies for application development. Practitioners could examine the significant challenges identified in this study to enhance their development approaches and practices. Furthermore, the research study emphasizes the importance of practitioners attaching accurate and appropriate tags to posted questions to facilitate effectively identifying challenges related to software development approaches. Several instances of posts inaccurately tagged with "development approaches" but asking specific technical questions were found, e.g., \textit{(What is the opposite of initialize (or init)?- SESE post \#163004)}. It is crucial to understand the definition of software development approaches. Therefore, future research could consider automated assistance to help practitioners correctly tag posted questions. Moreover, recognizing and mapping the challenges associated with different software development approaches can help practitioners make more informed decisions when selecting methodologies that best suit their projects, team dynamics, and organizational structure. This can lead to improved project outcomes and reduced risks of project failure.

%% file: threats.tex
\section{Threats to Validity}
\label{sec:threats}

The validity of this study may be affected by a range of potential threats. We have analyzed the possible threats in terms of the four fundamental types of validity threats, namely internal validity, external validity, construct validity, and conclusion validity, as outlined by Wohlin et al. \cite{wohlin2012experimentation}.
\subsection{Internal validity}
Internal validity refers to the degree to which certain factors influence the results and analysis of the extracted data. In the context of this study, potential threats to internal validity could occur during various phases, including:
\paragraph{Data collection limitations}
A potential threat to internal validity pertains to the data collection for the proposed approach. We gathered data on software development approaches from SO, SESE, and SEPM using tags assigned by software practitioners on these platforms. Our tag-based method required us to consider all related tags to enhance the proposed approach's performance. However, we limited our search to general tags such as “development approaches", “agile", “software development life cycle", and some commonly used software development framework names like “Scrum", “DevOps", and “Kanban". This approach may have missed some related practitioners' posts on software development approaches, although it effectively collected relevant developers' questions and excluded false positives. Guaranteeing 100\% relevance of over 13k collected posts without comprehensive manual validation is challenging. To minimize this threat, we excluded practitioners' posts containing code snippets and conducted analyses on the remaining data.
\paragraph{Topic modeling subjectivity}
Topic modeling with LDA has proven effective in analyzing large amounts of textual data. However, the method requires subjectivity in assigning labels to practitioners' topics based on individual understanding, which could lead to misinterpretation risks. To mitigate this threat, we manually reviewed 15 developers' posts with the highest relevance to each topic and cross-checked them with the paper's first three authors.

\subsection{External validity}
External validity refers to the degree to which the findings of a study can be generalized to other populations, settings, or conditions beyond the specific sample or context of the study. The generalizability of the proposed results may pose a threat to external validity. We performed a qualitative analysis of 200 highly ranked posts to complement the results from topic modeling with more in-depth insights. However, the findings of this qualitative analysis were based on a small sample of posts and may not be entirely generalizable. To address this limitation, we utilized AMS \citep{bajaj2014mining} to select developers' posts as possible representatives of practitioners' discussions on Q\&A platforms.

\subsection{Construct validity}
Construct validity refers to the degree to which the measures used in a study accurately measure the intended constructs or concepts of interest. In this study, a potential threat to construct validity is the operationalization of software development approaches based on tags and developers' posts. There might be discrepancies between the actual software development practices and the tags or posts analyzed in this study. To address this threat, we gathered data from multiple platforms (SO, SESE, and SEPM) and utilized a combination of topic modeling and qualitative analysis to provide a more comprehensive understanding of software development approaches.

\subsection{Conclusion validity}
Conclusion Validity pertains to the level of credibility or reasonableness of the study's conclusions. In this study, the conclusions may rely on the subjective interpretation and knowledge of a sole author, which could result in unaddressed disagreements or discrepancies between co-authors. To minimize this possibility, the first author conducted the gathering and analysis of study data. The rest of the authors thoroughly examined the data during several meetings. Any disparities or conflicts in data analysis were resolved through transparent discussions and cooperative efforts among all authors. Additionally, the study authors conducted several brainstorming sessions to arrive at the final conclusions.

%% file: conclusion.tex
\section{Conclusion and Future Work}
\label{sec:conclusion}

This study examined 13,903 software development approaches related to developers' posts on SO, SESE, and SEPM to understand current practices and trends in this domain from the software practitioners' perspective. We conducted topic modeling and qualitative analysis of the collected data. Although several research studies have investigated software development and hybrid approaches in the software industry \citep{kuhrmann2017hybrid, zhou2021system, aymerich2018software, mushashu2019investigating}, to the best of our knowledge, we conducted the first study to empirically explore practitioners' views on software development approaches by mining data from Q\&A sites. Compared to existing studies, our research findings offer exciting and valuable insights into various aspects of software development approaches from a broader perspective. 

The analysis of developers' posts revealed a gradual increase in the number of practitioners' questions about software development approaches. However, the trend of successful developers' questions concerning software development approaches has declined since 2014. We argue that software development approaches are continuously evolving, becoming more complex, advanced, and challenging for software developers to adopt. 

Furthermore, using the LDA algorithm, we identified 15 commonly discussed software development approach-related practitioner topics on Q\&A sites. Among these topics, we discovered popular (e.g., Software development methodology concepts (T10)) and difficult (e.g., Events, bounded contexts and Microservices in DDD (T2)) topics. Notably, all the top-3 most difficult topics experienced an upward trend over time. We also found a negative correlation (tau = -0.600, p-value = 0.001) between topic popularity and difficulty (popularity declined with increasing difficulty) using Kendall's Tau correlation test \citep{kendall1970rank}. Popular topics likely reflect the areas of interest and concern for practitioners and may represent the field's most critical concepts and practices.
These findings are significant regarding complex topics, as they highlight the need for greater attention and focus on improving practitioners' understanding and ability to apply them effectively.

We have also examined the challenges practitioners encounter when applying software development approaches. Our analysis identified 49 challenging factors that are further mapped across 14 sub-themes and 4 high-level themes.  Identifying and classifying these challenges is important because it signals that more attention and effort are required from experts and practitioners to address them adequately. Furthermore, mapping these challenging factors can enhance practitioners' comprehension of the issues in detail and help them better prepare for overcoming them while implementing software development approaches.

In the future, we plan to extend  this study by exploring additional challenging factors, identifying the causes of the challenges we have identified, and investigating best practices for addressing them. To achieve this, we intend to expand our study to include other well-known Q\&A repositories, such as GitHub, and conduct industrial surveys and interviews with software development approach experts to validate our findings and gather their insights.

%% file: main.bbl
\begin{thebibliography}{10}
\expandafter\ifx\csname url\endcsname\relax
  \def\url#1{\texttt{#1}}\fi
\expandafter\ifx\csname urlprefix\endcsname\relax\def\urlprefix{URL }\fi
\expandafter\ifx\csname href\endcsname\relax
  \def\href#1#2{#2} \def\path#1{#1}\fi

\bibitem{al2020agile}
S.~Al-Saqqa, S.~Sawalha, H.~AbdelNabi, Agile software development:
  Methodologies and trends., International Journal of Interactive Mobile
  Technologies 14~(11) (2020).

\bibitem{rico2009business}
D.~F. Rico, H.~H. Sayani, S.~Sone, The business value of agile software
  methods: maximizing ROI with just-in-time processes and documentation, J.
  Ross Publishing, 2009.

\bibitem{kuhrmann2017hybrid}
M.~Kuhrmann, P.~Diebold, J.~M{\"u}nch, P.~Tell, V.~Garousi, M.~Felderer,
  K.~Trektere, F.~McCaffery, O.~Linssen, E.~Hanser, et~al., Hybrid software and
  system development in practice: waterfall, scrum, and beyond, in: Proceedings
  of the 2017 international conference on software and system process, 2017,
  pp. 30--39.

\bibitem{kuhrmann2021makes}
M.~Kuhrmann, P.~Tell, R.~Hebig, J.~Kl{\"u}nder, J.~M{\"u}nch, O.~Linssen,
  D.~Pfahl, M.~Felderer, C.~R. Prause, S.~G. MacDonell, et~al., What makes
  agile software development agile?, IEEE transactions on software engineering
  48~(9) (2021) 3523--3539.

\bibitem{petersen2009waterfall}
K.~Petersen, C.~Wohlin, D.~Baca, The waterfall model in large-scale
  development, in: International Conference on Product-Focused Software Process
  Improvement, Springer, 2009, pp. 386--400.

\bibitem{khan2021agile}
A.~A. Khan, M.~Shameem, M.~Nadeem, M.~A. Akbar, Agile trends in chinese global
  software development industry: Fuzzy ahp based conceptual mapping, Applied
  Soft Computing 102 (2021) 107090.

\bibitem{kuhrmann2019walking}
M.~Kuhrmann, J.~Nakatumba-Nabende, R.-H. Pfeiffer, P.~Tell, J.~Kl{\"u}nder,
  T.~Conte, S.~G. MacDonell, R.~Hebig, Walking through the method zoo: does
  higher education really meet software industry demands?, in: 2019 IEEE/ACM
  41st International Conference on Software Engineering: Software Engineering
  Education and Training (ICSE-SEET), IEEE, 2019, pp. 1--11.

\bibitem{riaz2019implementation}
M.~N. Riaz, Implementation of kanban techniques in software development
  process: An empirical study based on benefits and challenges, Sukkur IBA
  Journal of Computing and Mathematical Sciences 3~(2) (2019) 25--36.

\bibitem{ali2020conceptualising}
J.~Ali~Khan, L.~Liu, L.~Wen, R.~Ali, Conceptualising, extracting and analysing
  requirements arguments in users' forums: The crowdre-arg framework, Journal
  of Software: Evolution and Process 32~(12) (2020) e2309.

\bibitem{khan2019analysis}
J.~A. Khan, Y.~Xie, L.~Liu, L.~Wen, Analysis of requirements-related arguments
  in user forums, in: 2019 IEEE 27th International Requirements Engineering
  Conference (RE), IEEE, 2019, pp. 63--74.

\bibitem{vidoni2022systematic}
M.~Vidoni, A systematic process for mining software repositories: Results from
  a systematic literature review, Information and Software Technology 144
  (2022) 106791.

\bibitem{zahedi2020mining}
M.~Zahedi, R.~N. Rajapakse, M.~A. Babar, Mining questions asked about
  continuous software engineering: A case study of stack overflow, in:
  Proceedings of the evaluation and assessment in software engineering, 2020,
  pp. 41--50.

\bibitem{haque2020challenges}
M.~U. Haque, L.~H. Iwaya, M.~A. Babar, Challenges in docker development: A
  large-scale study using stack overflow, in: Proceedings of the 14th ACM/IEEE
  International Symposium on Empirical Software Engineering and Measurement
  (ESEM), 2020, pp. 1--11.

\bibitem{le2021large}
T.~H.~M. Le, R.~Croft, D.~Hin, M.~A. Babar, A large-scale study of security
  vulnerability support on developer q\&a websites, in: Evaluation and
  assessment in software engineering, 2021, pp. 109--118.

\bibitem{zhou2020improving}
C.~Zhou, B.~Li, X.~Sun, Improving software bug-specific named entity
  recognition with deep neural network, Journal of Systems and Software 165
  (2020) 110572.

\bibitem{brisson2020we}
S.~Brisson, E.~Noei, K.~Lyons, We are family: analyzing communication in github
  software repositories and their forks, in: 2020 IEEE 27th International
  Conference on Software Analysis, Evolution and Reengineering (SANER), IEEE,
  2020, pp. 59--69.

\bibitem{paixao2017interplay}
K.~V. Paix{\~a}o, C.~Z. Fel{\'\i}cio, F.~M. Delfim, M.~d.~A. Maia, On the
  interplay between non-functional requirements and builds on continuous
  integration, in: 2017 IEEE/ACM 14th International Conference on Mining
  Software Repositories (MSR), IEEE, 2017, pp. 479--482.

\bibitem{dwivedi2018software}
A.~K. Dwivedi, A.~Tirkey, S.~K. Rath, Software design pattern mining using
  classification-based techniques, Frontiers of Computer Science 12~(5) (2018)
  908--922.

\bibitem{sun2015msr4sm}
X.~Sun, B.~Li, H.~Leung, B.~Li, Y.~Li, Msr4sm: Using topic models to
  effectively mining software repositories for software maintenance tasks,
  Information and Software Technology 66 (2015) 1--12.

\bibitem{blei2003latent}
D.~M. Blei, A.~Y. Ng, M.~I. Jordan, Latent dirichlet allocation, Journal of
  machine Learning research 3~(Jan) (2003) 993--1022.

\bibitem{kendall1938new}
M.~G. Kendall, A new measure of rank correlation, Biometrika 30~(1/2) (1938)
  81--93.

\bibitem{royce1987managing}
W.~W. Royce, Managing the development of large software systems: concepts and
  techniques, in: Proceedings of the 9th international conference on Software
  Engineering, 1987, pp. 328--338.

\bibitem{boehm1988spiral}
B.~W. Boehm, A spiral model of software development and enhancement, Computer
  21~(5) (1988) 61--72.

\bibitem{beck2001manifesto}
K.~Beck, M.~Beedle, A.~Van~Bennekum, A.~Cockburn, W.~Cunningham, M.~Fowler,
  J.~Grenning, J.~Highsmith, A.~Hunt, R.~Jeffries, et~al., Manifesto for agile
  software development, Snowbird, UT, 2001.

\bibitem{kim2021devops}
G.~Kim, J.~Humble, P.~Debois, J.~Willis, N.~Forsgren, The DevOps handbook: How
  to create world-class agility, reliability, \& security in technology
  organizations, IT Revolution, 2021.

\bibitem{sutherland2014scrum}
J.~Sutherland, J.~Sutherland, Scrum: the art of doing twice the work in half
  the time, Currency, 2014.

\bibitem{kuhrmann20172nd}
M.~Kuhrmann, P.~Diebold, S.~MacDonell, J.~M{\"u}nch, 2nd workshop on hybrid
  development approaches in software systems development, in: International
  Conference on Product-Focused Software Process Improvement, Springer, 2017,
  pp. 397--403.

\bibitem{fraser2007no}
S.~D. Fraser, F.~P. Brooks~Jr, M.~Fowler, R.~Lopez, A.~Namioka, L.~Northrop,
  D.~L. Parnas, D.~Thomas, No silver bullet" reloaded: retrospective on"
  essence and accidents of software engineering, in: Companion to the 22nd ACM
  SIGPLAN conference on Object-oriented programming systems and applications
  companion, 2007, pp. 1026--1030.

\bibitem{klunder2017helena}
J.~Kl{\"u}nder, P.~Hohl, M.~Fazal-Baqaie, S.~Krusche, S.~K{\"u}pper,
  O.~Linssen, C.~R. Prause, Helena study: Reasons for combining agile and
  traditional software development approaches in german companies, in:
  International Conference on Product-Focused Software Process Improvement,
  Springer, 2017, pp. 428--434.

\bibitem{zhou2021system}
P.~Zhou, A.~A. Ali~Khan, P.~Liang, S.~Badshah, System and software processes in
  practice: Insights from chinese industry, in: Evaluation and Assessment in
  Software Engineering, 2021, pp. 394--401.

\bibitem{aymerich2018software}
B.~Aymerich, I.~D{\'\i}az-Oreiro, J.~C. Guzm{\'a}n, G.~L{\'o}pez, D.~Garbanzo,
  Software development practices in costa rica: A survey, in: International
  Conference on Applied Human Factors and Ergonomics, Springer, 2018, pp.
  122--132.

\bibitem{mushashu2019investigating}
E.~T. Mushashu, J.~S. Mtebe, Investigating software development methodologies
  and practices in software industry in tanzania, in: 2019 IST-Africa Week
  Conference (IST-Africa), IEEE, 2019, pp. 1--11.

\bibitem{bajec2007practice}
M.~Bajec, D.~Vavpoti{\v{c}}, M.~Krisper, Practice-driven approach for creating
  project-specific software development methods, Information and Software
  technology 49~(4) (2007) 345--365.

\bibitem{bustard2013maturation}
D.~Bustard, G.~Wilkie, D.~Greer, The maturation of agile software development
  principles and practice: Observations on successive industrial studies in
  2010 and 2012, in: 2013 20th IEEE International Conference and Workshops on
  Engineering of Computer Based Systems (ECBS), IEEE, 2013, pp. 139--146.

\bibitem{tell2021towards}
P.~Tell, J.~Kl{\"u}nder, S.~K{\"u}pper, D.~Raffo, S.~MacDonell, J.~M{\"u}nch,
  D.~Pfahl, O.~Linssen, M.~Kuhrmann, Towards the statistical construction of
  hybrid development methods, Journal of Software: Evolution and Process 33~(1)
  (2021) e2315.

\bibitem{barua2014developers}
A.~Barua, S.~W. Thomas, A.~E. Hassan, What are developers talking about? an
  analysis of topics and trends in stack overflow, Empirical Software
  Engineering 19~(3) (2014) 619--654.

\bibitem{pinto2014mining}
G.~Pinto, F.~Castor, Y.~D. Liu, Mining questions about software energy
  consumption, in: Proceedings of the 11th Working Conference on Mining
  Software Repositories, 2014, pp. 22--31.

\bibitem{treude2011programmers}
C.~Treude, O.~Barzilay, M.-A. Storey, How do programmers ask and answer
  questions on the web?(nier track), in: Proceedings of the 33rd international
  conference on software engineering, 2011, pp. 804--807.

\bibitem{chen2016survey}
T.-H. Chen, S.~W. Thomas, A.~E. Hassan, A survey on the use of topic models
  when mining software repositories, Empirical Software Engineering 21~(5)
  (2016) 1843--1919.

\bibitem{ahmed2018concurrency}
S.~Ahmed, M.~Bagherzadeh, What do concurrency developers ask about? a
  large-scale study using stack overflow, in: Proceedings of the 12th ACM/IEEE
  international symposium on empirical software engineering and measurement,
  2018, pp. 1--10.

\bibitem{rosen2016mobile}
C.~Rosen, E.~Shihab, What are mobile developers asking about? a large scale
  study using stack overflow, Empirical Software Engineering 21~(3) (2016)
  1192--1223.

\bibitem{bajaj2014mining}
K.~Bajaj, K.~Pattabiraman, A.~Mesbah, Mining questions asked by web developers,
  in: Proceedings of the 11th Working conference on mining software
  repositories, 2014, pp. 112--121.

\bibitem{braun2006using}
V.~Braun, V.~Clarke, Using thematic analysis in psychology, Qualitative
  research in psychology 3~(2) (2006) 77--101.

\bibitem{cruzes2011recommended}
D.~S. Cruzes, T.~Dyba, Recommended steps for thematic synthesis in software
  engineering, in: 2011 international symposium on empirical software
  engineering and measurement, IEEE, 2011, pp. 275--284.

\bibitem{griffiths2004finding}
T.~L. Griffiths, M.~Steyvers, Finding scientific topics, Proceedings of the
  National academy of Sciences 101~(suppl\_1) (2004) 5228--5235.

\bibitem{jacobi2016quantitative}
C.~Jacobi, W.~Van~Atteveldt, K.~Welbers, Quantitative analysis of large amounts
  of journalistic texts using topic modelling, Digital journalism 4~(1) (2016)
  89--106.

\bibitem{khan2018linguistic}
J.~A. Khan, L.~Liu, Y.~Jia, L.~Wen, Linguistic analysis of crowd requirements:
  an experimental study, in: 2018 IEEE 7th International Workshop on Empirical
  Requirements Engineering (EmpiRE), IEEE, 2018, pp. 24--31.

\bibitem{yang2016security}
X.-L. Yang, D.~Lo, X.~Xia, Z.-Y. Wan, J.-L. Sun, What security questions do
  developers ask? a large-scale study of stack overflow posts, Journal of
  Computer Science and Technology 31~(5) (2016) 910--924.

\bibitem{abdellatif2020challenges}
A.~Abdellatif, D.~Costa, K.~Badran, R.~Abdalkareem, E.~Shihab, Challenges in
  chatbot development: A study of stack overflow posts, in: Proceedings of the
  17th international conference on mining software repositories, 2020, pp.
  174--185.

\bibitem{roder2015exploring}
M.~R{\"o}der, A.~Both, A.~Hinneburg, Exploring the space of topic coherence
  measures, in: Proceedings of the eighth ACM international conference on Web
  search and data mining, 2015, pp. 399--408.

\bibitem{peng2022topics}
A.~A. Khan, J.~A. Khan, M.~A. Akbar, P.~Zhou, M.~Fahmideh, Complete list of
  topics, top 10 words, top 15 posts and assigned labels,
  https://zhoupppp.notion.site/SPRPs-1da3b78e78234e5293becaaba846ef63,
  accessed: August 23, 2022 (2022).

\bibitem{kendall1970rank}
M.~Kendall, Rank correlation methods 4th edition charles griffin, High Wycombe,
  Bucks (1970).

\bibitem{wohlin2012experimentation}
C.~Wohlin, P.~Runeson, M.~H{\"o}st, M.~C. Ohlsson, B.~Regnell, A.~Wessl{\'e}n,
  Experimentation in software engineering, Springer Science \& Business Media,
  2012.

\end{thebibliography}
